\documentclass[11pt]{article}
\usepackage[utf8]{inputenc}
\usepackage{amsmath}
\usepackage{graphicx}
\usepackage{enumerate}
\usepackage{natbib}
\usepackage[english]{babel}
\usepackage{amsfonts}
\usepackage{amsthm}
\graphicspath{ {Figures/} }
\usepackage{setspace}
\usepackage[table, dvipsnames]{xcolor}
\newtheorem{theorem}{Theorem}
\newtheorem{corollary}{Corollary}

\newcommand{\blind}{1}

\begin{document}

\def\spacingset#1{\renewcommand{\baselinestretch}%
{#1}\small\normalsize} \spacingset{1}


\if1\blind
{\title{\bf $D$- and $A$-Optimal Screening Designs}
  \author{Jonathan Stallrich\\ 
  North Carolina State University, Department of Statistics\\   
  Katherine Allen-Moyer\\
    North Carolina State University, Department of Statistics\\ 
    Bradley Jones \\ JMP Statistical Discovery Software LLC}
  \maketitle
} \fi

\if0\blind
{
  \bigskip
  \bigskip
  \bigskip
  \begin{center}
    {\LARGE\bf $D$- and $A$-Optimal Screening Designs}
\end{center}
  \medskip
} \fi

\bigskip
\begin{abstract}
\noindent 
In practice, optimal screening designs for arbitrary run sizes are traditionally generated using the $D$-criterion with factor settings fixed at $\pm 1$, even when considering continuous factors with levels in $[-1,1]$. This paper identifies cases of undesirable estimation variance properties for such $D$-optimal designs and argues that generally $A$-optimal designs tend to push variances closer to their minimum possible value. New insights about the behavior of the criteria are found through a study of their respective coordinate-exchange formulas. The study confirms the existence of $D$-optimal designs comprised only of settings $\pm 1$ for both main effect and interaction models for blocked and unblocked experiments. Scenarios are also identified for which arbitrary manipulation of a coordinate between $[-1,1]$ leads to infinitely many $D$-optimal designs each having different variance properties. For the same conditions, the $A$-criterion is shown to have a unique optimal coordinate value for improvement. We also compare Bayesian version of the $A$- and $D$-criteria in how they balance minimization of estimation variance and bias. Multiple examples of screening designs are considered for various models under Bayesian and non-Bayesian versions of the $A$- and $D$-criteria.
\end{abstract}

\vfill

\newpage
\spacingset{1.45} 





\section{Introduction}\label{s:Intro}


A screening experiment is an initial step in a sequential experimental procedure to understand and/or optimize a process dependent upon many controllable factors. Such experiments are common in pharmaceuticals, agriculture, genetics, defense, and textiles (see \cite{dean2006screening} for a comprehensive overview of screening design methodology and applications).  The screening analysis aims to identify the few factors that drive most of the process variation often according to a linear model comprised of main effects, interaction effects, and, in the case of numeric factors, quadratic effects \citep{jones2011class}. Each effect corresponds to one or more factors, and a factor is said to be active if at least one of its corresponding effects is large relative to the process noise; otherwise the factor is said to be inert.  Analyses under this class of models often follow effect principles of sparsity, hierarchy, and heredity (see Chapter 9 of \cite{WuHamada}), with the primary goal of correctly classifying each factor as active or inert.  

A screening design is represented by an $n \times k$ matrix, $\boldsymbol{X}_d$, with rows $\boldsymbol{x}_i^T=(x_{i1},\dots,x_{ik})$ where $x_{ij}$ represents the $j$-th factor's setting for run $i$. 
To standardize screening designs across applications, continuous factor settings are scaled so $x_{ij} \in [-1,1]$ while categorical factor settings are often restricted to two levels, making $x_{ij}=\pm 1$.  We compare $\boldsymbol{X}_d$'s based on the statistical properties of the effects' least-squares estimators because their properties are tractable, particularly their variances and potential biases.  The goal then is to identify an $\boldsymbol{X}_d$ that minimizes the individual variances and biases of these effect estimators.


Suppose the model is correctly specified and there are designs having unique least-squares estimators for all effects. Then these estimators are unbiased and designs may be compared based on their estimation variances. A design having variances that are as small as possible will improve one's ability to correctly classify factors as active/inert. For models comprised solely of main effects and interactions, orthogonal designs have estimation variances simultaneously equal to their minimum possible value across all designs. Such designs exist only when $n$ is a multiple of 4; for other $n$ it is unclear which design will have the best variance properties. Still, designs should be compared based on how close their variances are to their respective minimum possible values. This approach requires knowledge of the minimum values as well as some measure of closeness. 

One approach for identifying minimum variances is to approximate them using the theoretical value assuming an orthogonal design exists, but such values may be unattainable. The $c$-criterion \citep{atkinson2007} may be used to identify the minimum variance for a given effect, but without any guarantee of the estimability of the other effects of interest. To remedy this estimability issue, \cite{AllenMoyer2021} proposed the $c_\mathcal{E}$-criterion to calculate these minimum variances exactly. It is less clear how to measure the proximity of a design's variances to their $c_\mathcal{E}$ values. The Pareto frontier approach by \cite{LuPareto2011} is well-suited for this problem but can be cumbersome in practice. A more practical solution is to evaluate and rank designs according to a single criterion that involves a scalar measure of all the variances. Such criteria should be straightforward to evaluate and optimize, and the resulting optimal designs should have variances close to their $c_\mathcal{E}$ values. Different forms of the $D$- and $A$-criterion (see Section~2.1) are popular variance-based criteria employed in the screening design literature and will be the focus of this paper. 







\begin{figure}[ht]
\begin{minipage}[b]{.48\textwidth}
\centering
\begin{tabular}{|rrrrr|}
\hline
    1 & 1 & 1 & 0 & 0 \\ 
   -1 & -1 & 1 & -1 & 1 \\ 
   -1 & 1 & -1 & -1 & 1 \\ 
    1 & -1 & -1 & -1 & -1 \\ 
   -1 & -1 & 1 & 1 & -1 \\ 
    1 & -1 & -1 & 1 & 1 \\ 
   -1 & 1 & -1 & 1 & -1 \\ 
     \hline
\end{tabular}
\end{minipage}
\hfill
\begin{minipage}[b]{.48\textwidth}
\centering
$\vcenter{\hbox{\includegraphics[width=.75\textwidth, angle = 270]{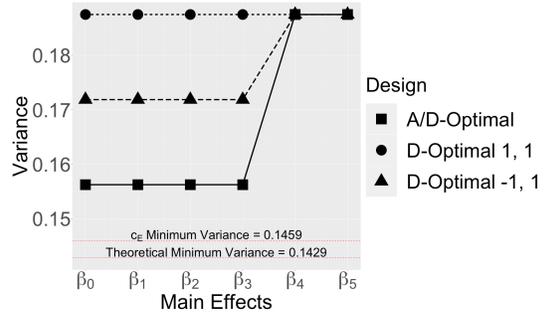}}}$
\end{minipage}
 \caption{(Left) $n = 7,\ k = 5$, $A$-optimal design. (Right) Main effect variances (in ascending order) for $A$- and $D$-optimal designs\label{tab:5F7Rex}. The design ``$D$-optimal $1, 1$'' replaces $x_{14}$ and $x_{15}$ of left design with $1$. Design ``D-optimal $-1, 1$'' is similarly defined. The minimum possible variances assuming an orthogonal design would each be $1/7=0.1429$ and the minimum variances under the $c_{\mathcal{E}}$-criterion from \cite{AllenMoyer2021} are $0.1459$.}
\end{figure}

Designs that optimize $D$- and $A$-criteria can coincide for some $n$, but this does not mean the criteria equivalently summarize variances. Consider a screening problem with $n=7$ runs and $k=5$ factors that assumes a main effect model. It is well-known that there always exists a $D$-optimal design comprised of $x_{ij}=\pm 1$, even when $x_{ij} \in [-1,1]$ \citep{box1971factorial}.  While other $D$-optimal designs having $x_{ij} \in (-1, 1)$ may exist, the screening literature predominantly fixes $x_{ij}=\pm 1$ with no assumed degradation to the resulting variances.  For example, \cite{jones2020Aoptimal} found an $A$-optimal design with $x_{ij}$ values of $\pm 1$ and $0$ having smaller variances compared to $D$-optimal designs comprised of $x_{ij}=\pm 1$ only. Figure~\ref{tab:5F7Rex} shows this $A$-optimal design, which has $x_{14}=x_{15}=0$.  Figure~\ref{tab:5F7Rex} also shows the corresponding main effect variances (in ascending order) of the $A$-optimal design and two $D$-optimal designs comprised of $x_{ij}=\pm 1$. The minimum possible variances assuming an orthogonal design exists are $1/7=0.1429$ and the minimum variances under the $c_{\mathcal{E}}$-criterion from \cite{AllenMoyer2021} are $0.1459$. Each of the $A$-optimal design's variances are equal to or smaller than the two competing $D$-optimal designs comprised of $\pm 1$.



As it turns out, the $A$-optimal design in Figure~1 is also $D$-optimal despite having some $x_{ij}=0$.  In fact, changing either $x_{14}$ or $x_{15}$ to any value in $[-1, 1]$ produces yet another $D$-optimal design but with equal or larger variances than the $A$-optimal design. The $A$-optimal design in this case, however, is unique. The existence of infinitely many $D$-optimal designs, each with equal or larger variances relative to the $A$-optimal design, is cause for concern about utilizing the $D$-criterion to rank screening designs. In this example, the $A$-criterion was better able to differentiate designs in terms of their ability to minimize the main effect variances simultaneously. 

This is not to say $D$-optimal designs are less valuable than $A$-optimal designs. Such designs have been used with great success in practice and the relative differences of the variances in Figure 1 do not appear large. Whether these differences impact the analysis depends on the ratio of the true main effect, denoted $\beta_j$, and the process variance, $\sigma^2$. When performing a two-sided $t$-test for the null hypothesis $\beta_j=0$, the associated noncentrality parameter will be $\beta_j/\sigma$ divided by the square root of the variances shown in Figure~1. When $\beta_j/\sigma$ is large, slight differences in the variances will not affect the noncentrality parameter, and hence will not affect power of the tests. The differences in variances will have a significant impact as $\beta_j/\sigma$ gets smaller. For example, suppose $\beta_j/\sigma=1$ and we perform a $t$-test for $\beta_1=0$ with significance level $\alpha=0.05$. The power for this test under the $D$-optimal design with $x_{14}=x_{15}=1$ is $0.6355$ while for the $A$-optimal design it is $0.7135$. Without any prior knowledge of the $\beta_j/\sigma$, it is important then to find a design that decreases the individual variances as much as possible.



Based on the effect principles, it is common to fit a main effect model even though interactions and/or quadratic effects may be active. The least-squares estimators for the main effect model may then become biased. Rather than try to estimate all potentially important effects, one can quantify the bias of the estimators and identify a design that simultaneously reduces estimation variance and bias. Let $\boldsymbol{\beta}$ be the vector of the largest collection of effects that may be important and hence captures the true model.  Partition $\boldsymbol{\beta}$ into $\boldsymbol{\beta}_1$ and $\boldsymbol{\beta}_2$ where $\boldsymbol{\beta}_1$ are effects we believe are most likely to be important and correspond to the effects in the fitted model, and $\boldsymbol{\beta}_2$ are the remaining effects that are potentially important but ignored in the fitted model. The possible bias from estimating $\boldsymbol{\beta}_1$ under the fitted model when the true model includes all $\boldsymbol{\beta}$ is $\boldsymbol{A}\boldsymbol{\beta}_2$ where $\boldsymbol{A}$ is the design's so-called alias matrix.  \cite{dumouchel1994simple} construct designs under model uncertainty by assigning a prior distribution to $\boldsymbol{\beta}_1$ and $\boldsymbol{\beta}_2$, and ranking designs according to the $D$-criterion applied to $\boldsymbol{\beta}$'s posterior covariance matrix.  While Bayesian $D$-optimal designs have shown an ability to balance minimizing bias and variance, the possible flaws of the $D$-criterion pointed out earlier are still concerning. Better designs may then be found with a Bayesian $A$-criterion, which has not received much attention in the screening literature.

This paper makes two important contributions that build a strong case for constructing screening designs under different forms of the $A$-criterion. The first contribution is a comparison of the behavior of the $D$- and $A$-criteria in response to manipulating a single coordinate of a given design. Our investigation not only provides insights into the criteria's coordinate exchange algorithms, a popular design construction algorithm, but also establishes the existence of $D$-optimal designs with $x_{ij}= \pm 1$ for models including main effects and/or interactions, as well as nuisance effects, such as block effects. We are only aware of such a result for main effect models with an intercept.  We also identify cases in which the $D$-criterion is invariant to any possible coordinate exchange, meaning the $D$-criterion considers all such designs as having equal value despite potentially having different variances. For such cases, we show that the $A$-criterion has a unique optimal coordinate exchange. Our second contribution is the promotion of a weighted Bayesian $A$-criterion for constructing designs that balance bias and variance minimization.  
We compare new screening designs generated under coordinate-exchange algorithms for common factorial models and show the Bayesian $A$-optimal designs have more appealing variance and bias properties than Bayesian $D$-optimal designs. 

The paper is organized as follows.  Section~\ref{s:ADReview} reviews traditional and current screening models and criteria. 
Section~\ref{s:Theory} investigates the behavior of the $D$- and $A$-criteria following coordinate exchanges to an existing design for models including nuisance effects. It also introduces the Bayesian $A$-criterion and how nuisance effects may be addressed under this criterion through a weight matrix.
Examples of $A$-optimal and Bayesian $A$-optimal designs constructed for main effect models, a two-factor interaction model, and a quadratic model are provided in Section~\ref{s:Bayes}. Section~\ref{s:block} constructs a blocked screening design for a pharmaceutical application under our new criteria.
We conclude the paper with a discussion of current and future work in Section~\ref{s:Discussion}.

\section{Background}\label{s:ADReview}

The fitted model for the $i$-th continuous response, $y_i$, has the form
\begin{equation}\label{eq:LinearModelVec}
    y_i = f^T(\boldsymbol{x}_i)\boldsymbol{\beta} + \boldsymbol{z}_i^T\boldsymbol{\theta} + e_i\ ,\  
\end{equation}
where $e_i \sim N(0,\sigma^2)$ and $i=1,\dots,n$.  Henceforth and without loss of generality, we set $\sigma^2 = 1$, since $\sigma^2$ is constant across all designs.  Every element of $f(\boldsymbol{x}_i)$, a $p \times 1$ vector, is a function of one or more elements of $\boldsymbol{x}_i$ while $\boldsymbol{z}_i$ is a $b \times 1$ vector that does not depend on $\boldsymbol{x}_i$ and corresponds to nuisance effects, $\boldsymbol{\theta}$. The simplest screening model is the main effect model where $f^T(\boldsymbol{x}_i) = \boldsymbol{x}^T_i$ and $z_i=1$, corresponding to an intercept effect, while a blocked main effect model with $b$ blocks has $\boldsymbol{z}_i$ comprised of all zeroes except for a $1$ in the $h$-th position when $y_i$ comes from block $h$.   Full quadratic models append the terms $\boldsymbol{x}^T_i \otimes \boldsymbol{x}^T_i=(x_{ij}x_{ij'})$ to the main effect model's $f^T(\boldsymbol{x}_i)$, where $\otimes$ denotes the Kronecker product.  Two-factor interaction models remove all $x_{ij}^2$ terms from the full quadratic model's $f(\boldsymbol{x}_i)$.  For a given $\boldsymbol{X}_d$, let $\boldsymbol{F}$ and $\boldsymbol{Z}$ denote matrices with rows $f(\boldsymbol{x}_i)$ and $\boldsymbol{z}_i$, respectively, and define $\boldsymbol{L}=(\boldsymbol{F}|\boldsymbol{Z})$.  







\subsection{Variance Criteria}

When model~\eqref{eq:LinearModelVec} is believed to contain the true model and $n > p+b$, we assume there exists at least one $\boldsymbol{X}_d$ with a unique least-squares estimator $(\hat{\boldsymbol{\beta}}^T|\hat{\boldsymbol{\theta}}^T)^T=(\boldsymbol{L}^T\boldsymbol{L})^{-1}\boldsymbol{L}^T\boldsymbol{y}$.  The estimator is unbiased and has variance $(\boldsymbol{L}^T\boldsymbol{L})^{-1}$.   Then $\text{Var}(\hat{\boldsymbol{\beta}})=\{\boldsymbol{F}^T(\boldsymbol{I}-\boldsymbol{P}_Z)\boldsymbol{F}\}^{-1}$ where $\boldsymbol{P}_Z=\boldsymbol{Z}(\boldsymbol{Z}^T\boldsymbol{Z})^{-1}\boldsymbol{Z}^T$ and screening inferences for the elements of $\hat{\boldsymbol{\beta}}$ perform best under an $\boldsymbol{X}_d$ whose $\text{Var}(\hat{\boldsymbol{\beta}})$ has small diagonal elements.  Designs may then be ranked based on a scalar function of
$(\boldsymbol{L}^T\boldsymbol{L})^{-1}$ that measure variance in some overall sense. To focus attention on estimation of $\boldsymbol{\beta}$, the function should be defined on $\text{Var}(\hat{\boldsymbol{\beta}})$.   All we require of $\boldsymbol{\theta}$ is that it can be estimated uniquely.



The $D$-criterion ranks designs according to $|(\boldsymbol{L}^T\boldsymbol{L})^{-1}|$ while the $D_s$-criterion is $|\text{Var}(\hat{\boldsymbol{\beta}})|$. In both cases, smaller values are desirable.  This paper uses the equivalent criteria of $|\boldsymbol{L}^T\boldsymbol{L}|$ and $|\boldsymbol{F}^T(\boldsymbol{I}-\boldsymbol{P}_Z)\boldsymbol{F}|$, with larger values being desirable. Under a normality assumption of $\boldsymbol{e}$, the $D$-optimal and $D_s$-optimal designs minimize the volume of the confidence ellipsoids for $(\boldsymbol{\beta}^T,\boldsymbol{\theta}^T)$ and $\boldsymbol{\beta}^T$, respectively. Hence these criteria are well-suited for an overall test of the significance of all effects, but not necessarily for individual testing of the parameters. The $A$-criterion ranks designs with respect to $\text{tr}[\{\boldsymbol{L}^T\boldsymbol{L}\}^{-1}]$ and the $A_s$-criterion is $\text{tr}[\text{Var}(\hat{\boldsymbol{\beta}})]$, being the sum of the individual variances of the parameters of interest. In both cases we want to minimize the chosen criterion.  

For main effect and interaction models, a design is said to be orthogonal when $\boldsymbol{L}^T\boldsymbol{L} = n\boldsymbol{I}$, meaning $\boldsymbol{F}$ is comprised of orthogonal columns of elements $\pm 1$ \citep{MukerjeeWu2006,WuHamada,Schoen2017}. 
Such designs estimate all main and interaction effects with the minimum possible variance, $1/n$.  By minimizing the individual variances, such designs will be both $D_s$- and $A_s$-optimal.  Orthogonal designs, however, can only exist when $n$ is a multiple of 4, otherwise the $D_s$- and $A_s$-optimal designs may differ from each other. Existing literature for constructing $A_s$- and $D_s$-optimal screening designs under arbitrary $n$ has predominantly focused on main effect models.  These designs are more commonly referred to as chemical balance and spring balance designs \citep{cheng1980optimality,masaro1983optimality,jacroux1983optimality,wong1984optimal,Cheng2014b}. To our knowledge, there are no theoretical results concerning $A_s$-optimal chemical balance designs with $x_{ij} \in \{\pm 1, 0\}$ for main effect models with an intercept nuisance effect.



\cite{jones2020Aoptimal} algorithmically constructed and compared $A$- and $D$-optimal designs under different screening models and arbitrary $n$.  They found that for $n = 3\ (\text{mod}\ 4)$ and $n$ small relative to $k$, $A$-optimal designs often had $x_{ij} \in \{\pm 1, 0\}$ .  In fact, they algorithmically constructed $A$-optimal designs allowing $x_{ij} \in [-1,1]$, yet still found the $A$-optimal designs only took on these three integer settings.  Similar to the $D$-optimal design's tendency to only have values $x_{ij}=\pm 1$, Sections~3.2 and 4.1 explore the conjecture that an $A$-optimal design exists where $x_{ij} \in \{\pm 1, 0\}$.  

\subsection{Variance and Bias Criteria}

While fitting the largest possible model incurs little to no bias, one needs a screening design with a  large run size ($n \geq p+b$).  
When $n < p+b$, there is no unique least-squares estimator and the analysis becomes more complicated.  Penalized regression, Bayesian methods, and stochastic model searches are increasing in popularity \citep{box1986analysis,Yuan2007,WuHamada,draguljic2014,Mee2017} and have proven to be quite powerful for screening.  
These analysis approaches, however, do not lend themselves to a tractable design framework. A design theory based on least-squares inference of submodels (e.g., \cite{Daniel1959}, \cite{lenth1989}, \cite{hamada1992analysis}, \cite{HamadaHamada2010}) is preferred. In practice, the main effects model should be the first submodel considered and subsequent models are chosen based on the results of that analysis according to the effect principles \citep{JonesNachtsheim2017}.

Partitioning $\boldsymbol{\beta}$ as in Section~1, a submodel may be thought of as fitting model~\eqref{eq:LinearModelVec} assuming $\boldsymbol{\beta}_2=0$.  Similarly partitioning $\boldsymbol{F}=(\boldsymbol{F}_1|\boldsymbol{F}_2)$ and defining $\boldsymbol{L}_1=(\boldsymbol{F}_1|\boldsymbol{Z})$, the least-squares estimator $(\hat{\boldsymbol{\beta}}_1^T|\hat{\boldsymbol{\theta}}^T)^T=(\boldsymbol{L}_1^T\boldsymbol{L}_1)^{-1}\boldsymbol{L}_1^T\boldsymbol{y}$.  
Fitting submodels introduces potential bias, namely the bias of $(\hat{\boldsymbol{\beta}}_1^T|\hat{\boldsymbol{\theta}}^T)^T$ is $\boldsymbol{A}\boldsymbol{\beta}_2$ where
\begin{align}
\boldsymbol{A}=(\boldsymbol{L}_1^T \boldsymbol{L}_1)^{-1}\boldsymbol{L}_1^T\boldsymbol{F}_2
\end{align}
is referred to as the alias matrix.  While variances for $\hat{\boldsymbol{\theta}}$ can be ignored when comparing designs, we should consider its bias with respect to $\boldsymbol{\beta}_2$ because we anticipate that some of these potential terms will be eventually considered in the analysis. The experimenter should then identify a design that minimizes both the diagonals of $\text{Var}(\hat{\boldsymbol{\beta}}_1)$ and the elements of $\boldsymbol{A}$.  


For $n = 0\ (\text{mod}\ 4)$, one strategy is to rank all strength 2 or 3 orthogonal arrays based on an aliasing-based criterion such as minimum aberration or one of its generalizations \citep{MukerjeeWu2006, Cheng2014,Vazquex2022}.  Doing so guarantees minimum variances after fitting the main effect model with minimal bias due to model misspecification.  For arbitrary $n$, \cite{jones2011efficient} and \cite{LuPareto2011} algorithmically optimize criteria that are some combination of the $D$-criterion and $\text{tr}[\boldsymbol{A}^T\boldsymbol{A}]$ under a given partition of $\boldsymbol{\beta}$. Bias may also be reduced through one's ability to fit many possible submodels, which is the goal of estimation capacity and model robust designs \citep{Li2000,Chen2004,Tsai2010,Smucker2012}, but such criteria are computationally intensive to calculate.


\cite{dumouchel1994simple} proposed a flexible Bayesian $D$-criterion to balance main effect variance and bias minimization. A uniform, improper prior is assigned to $\boldsymbol{\beta}_1$ and $\boldsymbol{\theta}$, and a $N(\boldsymbol{0}, \tau^2\boldsymbol{I}_q)$ prior to $\boldsymbol{\beta}_2$.  For $\boldsymbol{y} \mid \boldsymbol{\beta},\boldsymbol{\theta}\sim N(\boldsymbol{F}\boldsymbol{\beta}+\boldsymbol{Z}\boldsymbol{\theta},\ \boldsymbol{I})$, the posterior covariance matrix for $(\boldsymbol{\beta}^T,\boldsymbol{\theta}^T)^T$ is then $(\boldsymbol{L}^T\boldsymbol{L} + \tau^{-2}\boldsymbol{K})^{-1}$ where $\boldsymbol{K}$ is a diagonal matrix with $0$'s for the corresponding $p$ primary terms and $1$ for the corresponding $q$ potential terms. The Bayesian $D$-criterion is $|\boldsymbol{L}^T\boldsymbol{L} + \tau^{-2}\boldsymbol{K}|$,
where $\tau^{-2}$ tunes the importance of minimizing bias and/or estimation of the potential terms.  As $\tau^{-2} \to \infty$, the criterion will be less influenced by changes in aliasing between the primary and potential terms since $\tau^{-2}\boldsymbol{I}_q$ will have large diagonal elements.  As $\tau^{-2} \to 0$, the potential terms become primary terms. \cite{dumouchel1994simple} recommended $\tau^{-2} = 1$ 
and constructed optimal designs via a coordinate exchange algorithm.  Other Bayesian approaches have been considered \citep{Toman1994,Joseph2006,Bingham2007,TSAI2007619} but with only only two or three level factors.  This paper will also explore the Bayesian $A$-criterion, $\text{tr}[(\boldsymbol{L}^T\boldsymbol{L} + \tau^{-2}\boldsymbol{K})^{-1}]$, and Bayesian $A_s$-criterion, being the trace of the submatrix of  $(\boldsymbol{L}^T\boldsymbol{L} + \tau^{-2}\boldsymbol{K})^{-1}$ corresponding to $\boldsymbol{\beta}$.

\section{Properties of the $D$- and $A$-criterion}\label{s:Theory}

It is challenging to analytically derive optimal designs for a given criterion under an arbitrary $n$ and $k$.  In practice, these criteria are optimized via some computer search algorithm, such as the coordinate exchange algorithm \citep{meyer1995coordinate}, branch-and-bound algorithms \citep{Ahipasaoglu2021}, and nonlinear programming \citep{EstebanBravo2017,Duarte2020}. While the two latter algorithms offer some guarantees of identifying the true optimum, the coordinate exchange algorithm is straightforward to implement and is employed in popular statistical software.  We focus on the coordinate exchange algorithm (CEA) in this section not only because of its wide adoption, but because it provides an analytical tool to study the behavior of these different forms of the $D$- and $A$-criterion defined in Section~2.

Let $\mathcal{X}_j$ denote the set of permissible coordinates for $x_{ij}$, making $\mathcal{X}= \mathcal{X}_1 \times \dots \times \mathcal{X}_k$ the set of permissible rows for $\boldsymbol{X}_d$.  Then $\mathcal{X}_j=\pm 1$ for categorical factors and $\mathcal{X}_j = [-1,1]$ for numeric factors.  A row exchange of an initial design, $\boldsymbol{X}_{d0}$, exchanges one of its existing rows, $\boldsymbol{x}_i$, with a candidate row $\tilde{\boldsymbol{x}} \in \mathcal{X}$. This leads to a row exchange of $f(\boldsymbol{x}_i)$, the $i$-th row of the initial design's model matrix, $\boldsymbol{F}_0$, with the candidate model matrix row, $f(\tilde{\boldsymbol{x}})$. No exchange is made to $\boldsymbol{Z}$, since nuisance effects are not design dependent. Hence an exchange gives a new design and model matrix, denoted $\widetilde{\boldsymbol{X}}$ and $\widetilde{\boldsymbol{L}}$, respectively. A row exchange algorithm (REA) for a given criterion identifies the optimal $\tilde{\boldsymbol{x}}$ for each $\boldsymbol{x}_i$ sequentially, updating $\boldsymbol{X}_{d0}$ one row at a time.  After going through all $n$ runs, the algorithm starts again at $\boldsymbol{x}_1$.  The process repeats itself until some convergence criterion is met.  The algorithm is often repeated across many initial starting designs and the overall best design is reported.  The reported design is not guaranteed to be globally optimal, but it is common in the screening literature to refer to them as optimal. More details about REAs may be found in \cite{atkinson2007}.

A coordinate exchange is a specific row exchange that only manipulates $x_{ij}$. Then we may partition $\boldsymbol{x}_i^T=(x_{ij} | \boldsymbol{x}_{i, -j})$ and represent the $i$-th row of $\boldsymbol{L}$ as
\begin{align}
l(\boldsymbol{x}_i) = \begin{pmatrix}
f_1(x_{ij})  \\ \hline
f_2(\boldsymbol{x}_{i, -j})\\
\boldsymbol{z}_i
\end{pmatrix}=\begin{pmatrix}l_1(x_{ij})\\ \hline l_2(\boldsymbol{x}_{i, -j})
\end{pmatrix}\ ,\ \label{e:partition}
\end{align}
where $f_1(x_{ij})=l_1(x_{ij})$ is the subvector of $f(\boldsymbol{x}_i)$ that only involves $x_{ij}$ and $f_2(\boldsymbol{x}_{i, -j})$
are the remaining elements.  For example, exchanging $x_{i1}$ for a two-factor interaction model with an intercept nuisance parameter has $l_1^T(x_{i1}) = (x_{i1}, \ x_{i1}x_{i2},\dots, \ x_{i1}x_{ik})$ and $l_2^T(\boldsymbol{x}_{i, -j})=( x_{i2},\dots, \ x_{ik},\ x_{i2}x_{i3},\dots, \ x_{i(k-1)}x_{ik},1)$. 
\cite{meyer1995coordinate} proposed the CEA that proceeds in the same fashion as a REA, but for a given $\boldsymbol{x}_i$, each coordinate $x_{ij}$ is updated sequentially. As the number of candidate coordinates $\mathcal{X}_j$ is smaller than $\mathcal{X}$, a CEA involves fewer computations and does not require the user to specify all possible candidate points in $\mathcal{X}$. Moreover, there exist fast update formulas for the forms of the $D$- and $A$-criterion considered in this paper that do not require repeated matrix inversion. However, compared to a REA, a CEA requires more random starts to avoid converging to a local optimum. The remainder of this section investigates the behavior of the CEA for the different forms of the $D$- and $A$-criteria.





\subsection{Properties of $D$-criterion}\label{s:subDoptTheory}

For an $\boldsymbol{x}_i$ in $\boldsymbol{X}_{d0}$, the $D$-criterion's REA seeks 
the exchange $\tilde{\boldsymbol{x}}$ that maximizes
\begin{align}
    \Delta_D(\boldsymbol{x}_i, \tilde{\boldsymbol{x}}) &= \frac{|\widetilde{\boldsymbol{L}}^T\widetilde{\boldsymbol{L}}|}
    {|\boldsymbol{L}^T_{0}
    \boldsymbol{L}_{0}|}
    = l^T(\tilde{\boldsymbol{x}})\boldsymbol{V}\, l(\tilde{\boldsymbol{x}}) + \{1 - v(\boldsymbol{x}_i)\}\label{eq:Dobj_REA}
\end{align}
where  $\boldsymbol{V} = \{1 - v(\boldsymbol{x}_i)\}\boldsymbol{D} + \boldsymbol{D} l(\boldsymbol{x}_i)\,  l^T(\boldsymbol{x}_i) \boldsymbol{D}$ with $\boldsymbol{D}=(\boldsymbol{L}^T_{0}
    \boldsymbol{L}_{0})^{-1}$and $v(\boldsymbol{x}_i)=l(\boldsymbol{x}_i)^T\boldsymbol{D}\, l(\boldsymbol{x}_i)$. The matrix $\boldsymbol{V}$ is symmetric and it is positive definite if and only if $v(\boldsymbol{x}_i) < 1$. If $v(\boldsymbol{x}_i)=1$ then $\boldsymbol{V}$ is positive semidefinite.  
For a coordinate exchange of $x_{ij}$ for $\tilde{x}$, we can permute the rows and columns of $\boldsymbol{V}$ following \eqref{e:partition} giving a function with respect to $\tilde{x}$
\begin{equation}\label{eq:Dobjfun}
   \Delta^{ij}_D(\tilde{x}) = l^T_1(\tilde{x})\boldsymbol{V}_{11} l_1(\tilde{x}) + \boldsymbol{a}^Tl_1(\tilde{x}) + c 
\end{equation}
\noindent where $\boldsymbol{a} = 2\boldsymbol{V}_{12}l_2(\boldsymbol{x}_{i, -j})$ and $c =l^T_2(\boldsymbol{x}_{i, -j})\boldsymbol{V}_{22}l_2(\boldsymbol{x}_{i, -j}) + \{1 - v(\boldsymbol{x}_i)\}$ are fixed.  

The CEA for the $D_s$-criterion can be done equivalently through the CEA for the $D$-criterion because  $|\boldsymbol{L}^T\boldsymbol{L}|=|\boldsymbol{Z}^T\boldsymbol{Z}| \times |\boldsymbol{F}^T(\boldsymbol{I}-\boldsymbol{P}_Z)\boldsymbol{F}|$.
That is, $\Delta^{ij}_D$ evaluates the ratio $|\widetilde{\boldsymbol{F}}^T(\boldsymbol{I}-\boldsymbol{P}_Z)\widetilde{\boldsymbol{F}}|/|\boldsymbol{F}_0^T(\boldsymbol{I}-\boldsymbol{P}_Z)\boldsymbol{F}_0|$, corresponding to the $D_s$-criterion. The CEA for the Bayesian $D$-criterion has a similar update formula to \eqref{eq:Dobjfun} but with matrix $\boldsymbol{D}=(\boldsymbol{L}_0^T\boldsymbol{L}_0 + \tau^{-2}\boldsymbol{K})^{-1}$  \citep{dumouchel1994simple}. The CEA for the Bayesian $D_s$-criterion is easily shown to be equivalent to the Bayesian $D$-criterion, similar to the equivalence of the CEAs for the $D$- and $D_s$-criterion. We refer to the collection of these four criteria (i.e., $D$, $D_s$, Bayesian $D$, and Bayesian $D_s$) as the $\mathcal{D}$-criteria.






We now provide a general result about optimal designs for the $\mathcal{D}$-criteria under what we call an $m$-factor interaction model. Let $J_m=\{j_1,\dots,j_m\}$ be a subset of $m$ of the $k$ factor indices. An $m$-factor interaction model has elements of $f(\boldsymbol{x})$ comprised only of
\begin{itemize}
    \item all $k$ main effect coordinates $(x_j)$;
    \item at least one coordinate of the form  $\prod_{j \in J_m} x_j$ for some $J_m$;
    \item any remaining coordinates are of the form $\prod_{j \in J} x_j$ where $|J|=2,\dots,m$.
\end{itemize}
The main effect model is then the one-factor interaction model. Equation~\eqref{eq:Dobjfun} provides a proof technique 
for the following theorem:
\begin{theorem}\label{Thm:Dlevels}
For any $m$-factor interaction model where $\mathcal{X}_j = \pm 1$ or $\mathcal{X}_j \in [-1,1]$, there exists an optimal design comprised of $x_{ij}=\pm 1$ for each of the $\mathcal{D}$-criteria.
\end{theorem}
\noindent This proof and all subsequent proofs are provided in the Supplementary Materials.  To our knowledge this result has been proven only for main effect models \citep{box1971factorial, mitchell1974algorithm, galil1980d} and non-Bayesian criteria.  Not only does our result extend to $m$-factor interaction models, it also applies to any nuisance effect structure that is design independent.

A practical consequence of Theorem~\ref{Thm:Dlevels} is that to algorithmically construct an optimal design for such models under one of the $\mathcal{D}$-criteria, we can restrict $\mathcal{X}_{j}=\pm 1$. An unfortunate consequence of Theorem~1 that highlights a potential deficiency is the following corollary: 
\begin{corollary}\label{Cor:Dlevels}
For any $m$-factor interaction model, suppose there exists an optimal design with respect to one of the $\mathcal{D}$-criteria where at least one $x_{ij}\neq \pm 1$ where $\mathcal{X}_j \in [-1,1]$. Then $\Delta_D^{ij}$ for that criterion is constant with respect to $\tilde{x}$, making all such possible exchanges produce another optimal design.
\end{corollary}
\noindent The phenomenon described in Corollary~\ref{Cor:Dlevels} occurred in Figure~1 for both coordinates $x_{14}$ and $x_{15}$ under the $D$- and $D_s$-criterion.  
Indeed, the designs with $(x_{i4},x_{i5})=(\pm 1,\pm1)$ produced the worst individual main effect variances.  This example raises doubts about the $\mathcal{D}$-criteria's ability to evaluate a design's screening abilities.

\subsection{ Properties of $A$-criterion}\label{s:subAoptTheory}
The decrease in the $A$-criterion following a row exchange is
\begin{align}\label{eq:RowExchange}
    \Delta_A(\boldsymbol{x}_i, \tilde{\boldsymbol{x}}) &= \text{tr}\{(\boldsymbol{L}_0^T\boldsymbol{L}_0)^{-1}\} - \text{tr}\{(\widetilde{\boldsymbol{L}}^T\widetilde{\boldsymbol{L}})^{-1}\}
        = \frac{l^T(\tilde{\boldsymbol{x}})\boldsymbol{U}l(\tilde{\boldsymbol{x}}) - \phi(\boldsymbol{x}_i)}{\Delta_D(\boldsymbol{x}_i, \tilde{\boldsymbol{x}})}
\end{align}
where $\boldsymbol{U} = \boldsymbol{V}\boldsymbol{D} + \boldsymbol{D}\boldsymbol{V} - [\{1 - v(\boldsymbol{x}_i)\}\boldsymbol{D} + \phi(\boldsymbol{x}_i)\boldsymbol{I}]\boldsymbol{D}$ and $\phi(\boldsymbol{x}_i)=l^T(\boldsymbol{x}_i)\boldsymbol{D}\boldsymbol{D}l(\boldsymbol{x}_i)$.  
The optimal exchange maximizes \eqref{eq:RowExchange}.  Unlike with $\boldsymbol{V}$, 
$l^T(\tilde{\boldsymbol{x}})\boldsymbol{U}l(\tilde{\boldsymbol{x}})$ can take on positive and negative values.
Partitioning $\boldsymbol{U}$ as we did with $\boldsymbol{V}$, the coordinate objective function is
\begin{equation}\label{eq:DeltaL}
\Delta_A^{ij}(\tilde{x}) = \frac{l^T_1(\tilde{x})\boldsymbol{U}_{11} l_1(\tilde{x}) + \boldsymbol{b}^Tl_1(\tilde{x}) + d }{\Delta_D^{ij}(\tilde{x})}\ ,\
\end{equation}
where $\boldsymbol{b} = 2\boldsymbol{U}_{12}l_2(\boldsymbol{x}_{i, -j})$ and $d = l^T_2(\boldsymbol{x}_{i, -j})\boldsymbol{U}_{22}l_2(\boldsymbol{x}_{i, -j}) - \phi(\boldsymbol{x}_i)$ are constant. 

The equivalence between the $D$- and $D_s$-criterion does not hold for the $A$- and $A_s$-criterion. Other than special cases \citep{Nachtsheim1989} there is no closed-form coordinate exchange formula for $A_s$. Computing the update after a row/coordinate exchange may be accomplished by first updating $(\boldsymbol{L}_0^T\boldsymbol{L})_0^{-1}$ via the Sherman-Morrison-Woodbury formula \citep{sherman1950adjustment} and directly calculating the change, denoted $\Delta_{A_s}^{ij}$. This will not be as computationally efficient as evaluating \eqref{eq:DeltaL}. 
Following \cite{StallingsMorgan2015}, let $\boldsymbol{W}$ be a diagonal matrix of $p+b$ elements where the first $p$ diagonal entries corresponding to $\boldsymbol{\beta}$ equal 1 and the last $b$ elements corresponding to $\boldsymbol{\theta}$ equal an arbitrarily small value, $w>0$.  The weighted $A$-criterion, or $A_W$-criterion, is then
\begin{align}
    \text{tr}[\boldsymbol{W}(\boldsymbol{L}^T\boldsymbol{L})^{-1}]=\sum_j \text{Var}(\hat{\beta}_j) + w \sum_h \text{Var}(\hat{\theta}_h)\ .\ \label{e:Anuisance2}
\end{align}
The coordinate exchange update for the $A_W$-criterion, denoted $\Delta_{AW}^{ij}$, is similar to \eqref{eq:DeltaL} and is derived in the Supplementary Materials.  Note the $A$-criterion is a special case of the $A_W$-criterion with $\boldsymbol{W}=\boldsymbol{I}$. 

From \eqref{e:Anuisance2}, we see $\lim_{w\to 0}\text{tr}[\boldsymbol{W}(\boldsymbol{L}^T\boldsymbol{L})^{-1}]=\text{tr}\left[\{\boldsymbol{F}^T(\boldsymbol{I}-\boldsymbol{P}_Z)\boldsymbol{F}\}^{-1}\right]$, the $A_s$-criterion. Therefore, $\lim_{w\to0} \Delta_{AW}^{ij} = \Delta_{A_s}^{ij}$. This result provides an efficient way to perform a CEA for the $A_s$-criterion using the $A_W$-criterion and setting $w$ to an arbitrarily small value. We have found $w=10^{-6}$ to perform well for most applications. This limiting result also allows us to study the behavior of $\Delta_{A_s}^{ij}$ through the more tractable $\Delta_{AW}^{ij}$.

The update formula for a coordinate exchange under the Bayesian $A$-criterion takes the same form as \eqref{eq:DeltaL} but with $\boldsymbol{D}=(\boldsymbol{L}_0^T\boldsymbol{L}_0 + \tau^{-2}\boldsymbol{K})^{-1}$. For the Bayesian $A_s$-criterion, we can apply the weighted approach to the posterior covariance matrix,
\begin{align}
    \text{tr}[\boldsymbol{W}(\boldsymbol{L}^T\boldsymbol{L}+\tau^{-2}\boldsymbol{K})^{-1}]\ .\ \label{e:BayesAnuisance}
\end{align}
We refer to this as the Bayesian $A_W$-criterion. To our knowledge, this is one of the earliest attempts at combining techniques from the weighted and Bayesian optimality literature. The Bayesian $A_W$- and Bayesian $A_s$-criterion's ability to balance minimization of the primary variances and their aliasing with the potential terms is investigated with multiple examples in Section~4. We collectively refer to the different criteria discussed here as the $\mathcal{A}$-criteria.








We initially sought to prove the conjecture that for any of the $\mathcal{A}$-criteria there always exists an optimal design such that all $x_{ij} \in \{\pm 1, 0\}$ for $m$-factor interaction models.  For such models and criteria, the coordinate update formula is a ratio of two quadratic polynomials with respect to $\tilde{x}$ and the optimum coordinate exchange can be found using fractional programming methods \citep{dinkelbach1967nonlinear}.  In the Supplementary Materials, we identify situations where the optimum is unique and occurs at a non-integer value.  This result by itself does not disprove the conjecture, but it does provide evidence to the contrary.  Section~4.1 further explores this conjecture algorithmically for certain $n$ and $k$ under the main effect model.  

We next considered the unfortunate scenario in Corollary~1 with respect to the $\mathcal{A}$-criteria. As demonstrated in \eqref{eq:DeltaL}, the coordinate update formula for each $\mathcal{A}$-criteria involves a coordinate update for some $\mathcal{D}$-criteria. 
\begin{corollary}\label{lem:Aconstant1} For an $m$-factor interaction model and design, $\boldsymbol{X}_{d0}$, consider one of the weighted criteria among the $\mathcal{A}$-criteria for $w>0$. If the update formula for the corresponding among the $\mathcal{D}$-criteria is constant, then $\Delta_{AW}^{ij}$ is uniquely maximized.  Moreover, $\Delta_{A_s}^{ij}$ is uniquely maximized when
$\boldsymbol{z}_i^T(\boldsymbol{Z}^T\boldsymbol{Z})^{-1}\boldsymbol{z}_i < 1$.
\end{corollary}
\noindent 
The corollary's condition $\boldsymbol{z}_i^T(\boldsymbol{Z}^T\boldsymbol{Z})^{-1}\boldsymbol{z}_i < 1$ holds for practical cases of an intercept-only nuisance effect and block effects from $b$ blocks each of size $2$ or more. It provides further support for the $\mathcal{A}$-criteria's ability to better differentiate designs than the $\mathcal{D}$-criteria.  

\section{Examples}\label{s:Bayes}
This section compares properties of algorithmically-generated optimal designs for three common screening models: (1) main effect models, (2) two-factor interaction models, and (3) quadratic models.  All models have an intercept-only nuisance effect. For main effect models, we utilize the $A_s$- and $D_s$-criterion. For the models, we also consider their Bayesian versions. The best designs generated are compared in terms of their main effect variances after fitting the main effect submodel and, when applicable, their aliasing with potential terms (two-factor interactions and/or quadratic effects).

\subsection{Main Effect Models}\label{s:CoordExchange}


A main effect model with $k$ factors has the scalar form
\[
y_{i}=\beta_0 + \sum_{j=1}^k x_{ij} \beta_j + e_i\ ,\
\]
where $\beta_0$ is an intercept and $\beta_j$ with $j>0$ are the main effects. We constructed $A_s$- and $D_s$-optimal designs under this model for $k = 3, \dots, 20$ factors and $n = k+1,\dots,24$ runs assuming either only discrete settings ($\mathcal{X}_j = \{\pm 1, 0\}$) or only continuous settings ($\mathcal{X}_j = [-1,1]$).  For continuous settings, we optimized \eqref{eq:DeltaL} with box-constrained L-BFGS over $[-1,1]$ \citep{byrd1995limited}.  Due to the demanding computations involved as $n$ and $k$ increase, each CEA was first performed with 100 random starts for both the continuous and discrete CEAs.  We then recorded the best criterion value for the algorithms separately and the overall best value.  If the two values were equal, we declared the value as optimal.  Otherwise, the CEA with the inferior value was performed again with another 100 random starts.  If the best value among this new batch did not improve the previous overall best value, the search was stopped.  If the value did improve the overall best value, the other CEA was run for another 100 starts and the iterative process continued.  For our investigation, this back-and-forth search took no more than 1000 overall total starting designs.  The $D_s$- or $A_s$-optimal designs were the designs with the best $D_s$- or $A_s$-value found across all iterations of both the discrete and continuous CEAs.

\begin{figure}[ht]
\centering
\includegraphics[width=1\textwidth]{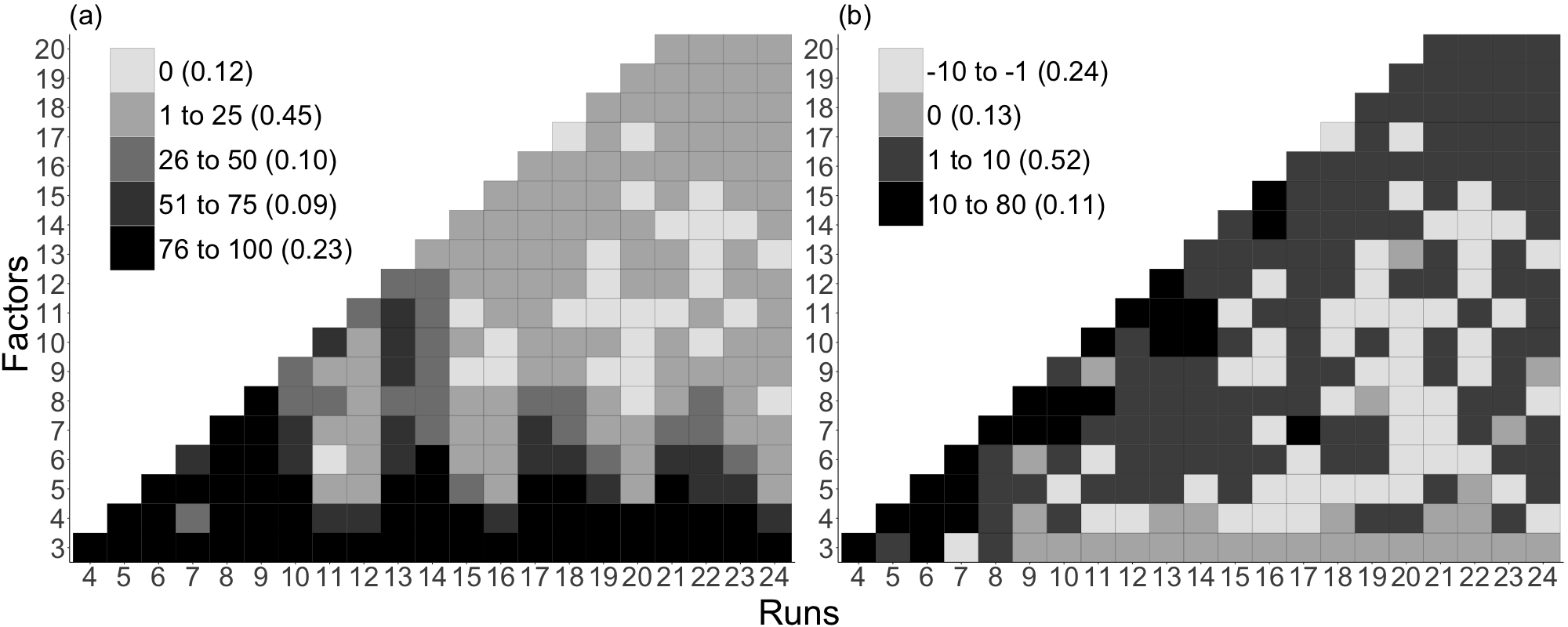}
\caption{(a) Categories of the number of times out of intial 100 starting designs the continuous CEA identified an $A_s$-optimal design for given $(n,k)$ combinations.  The overall proportion for each category is shown in parentheses. (b) Difference between the number of times the initial starting designs for the continuous and discrete CEA identified an $A_s$-optimal design for given $(n,k)$ combinations. \label{fig:CoordHeatMap}}
\end{figure}

Figure~\ref{fig:CoordHeatMap}(a) shows how many of the initial 100 constructed designs under the continuous CEA were $A$-optimal. A 0 value means the continuous CEA never found an $A_s$-optimal design among the initial batch of 100 random starting designs. Figure~\ref{fig:CoordHeatMap}(b) shows the difference between the counts in Figure~\ref{fig:CoordHeatMap}(a) with the same counts under the discrete CEA.  Generally, when $n = k + 1$ or $k + 2$, the continuous CEA identified an $A_s$-optimal design more frequently than the discrete CEA.  The discrete CEA found the $A_s$-optimal design more frequently in only 24\% of the scenarios considered and struggled particularly in the cases of $(n, k)=(11, 10)$ and $(19, 18)$. For these cases, even increasing the number of starting designs to $10,000$, the discrete CEA was unable to find an $A_s$-optimal design.  The continuous CEA was able to find an $A_s$-optimal design for all cases when we increased the number of starting designs to $1000$.   We therefore recommend the continuous CEA for constructing $A_s$-optimal designs.


Contrary to our conjecture in Section~3.2, the $A_s$-optimal designs found by the continuous CEA for scenarios $(11, 10)$ and $(19, 18)$ contained non-integer values and are displayed in the Supplementary Materials. These designs, however, do not significantly decrease the $A_s$-criterion compared to the best constructed designs requiring $\mathcal{X}_j=\pm 1$ or $\{\pm 1, 0\}$ as given in \cite{jones2020Aoptimal}. The criterion value for the $(11, 10)$ $A_s$-optimal design we constructed was only $0.28\%$ and $0.35\%$ more $A_s$-efficient than the three- and two-level designs, respectively. The efficiency for the $(19, 18)$ $A_s$-optimal design with non-integer factor settings was $0.08\%$ than the best discrete-level $A_s$-optimal design we generated. 

Similar to \cite{jones2020Aoptimal}, the main effect variances the $A_s$- and $D_s$-optimal designs we generated were the same except when $n$ was close to $k$ or when $n = 3\ (\text{mod}\ 4)$. For the designs where $n = 3\ (\text{mod}\ 4)$, we calculated the paired differences between the ordered main effect variances of the two designs. Across all such scenarios and pairs, $78\%$ of the main effect variances from the $A_s$-optimal designs were smaller than those from $D_s$-optimal designs, $6\%$ of them were equal, and $16\%$ of them had larger variances for the $A_s$-optimal design. The largest individual decrease an $A_s$-optimal design's variance had compared to the $D_s$-optimal design was $0.05$. There was one scenario where the $D_s$-optimal design decreased a variance over the $A_s$-optimal design by $0.06$.  

\subsection{Two-factor Interaction Model with $n = 15, \ k = 6$}\label{subsec:2fibayes}
Under the main effect model for scenario $n=15$, $k=6$, the $A_s$- and $D_s$-optimal designs were different, with the $A_s$-optimal design having zero coordinates for factors $5$ and $6$. We now consider this scenario under a two-factor interaction model: 
\[
y_{i}=\beta_0 + \sum_{j=1}^k x_{ij} \beta_j + \sum_{1 \leq j<j' \leq k} x_{ij}x_{ij'} \beta_{jj'}+ e_i\ ,\
\]
which adds $k(k-1)/2$ interaction effects $\beta_{jj'}$. Not all effects can be estimated uniquely due to the small $n$. Thus we constructed Bayesian $A_s$- and Bayesian $D_s$-optimal designs where the intercept is a nuisance effect, main effects are primary terms, and two-factor interaction effects are potential terms. We set $\tau^{-2} = 1,5, 10, \dots,100$ and for each value we performed a CEA with $1000$ starting designs. 

\begin{figure}
\begin{minipage}[b]{.45\textwidth}
\centering
$\vcenter{\hbox{\includegraphics[width=1\textwidth]{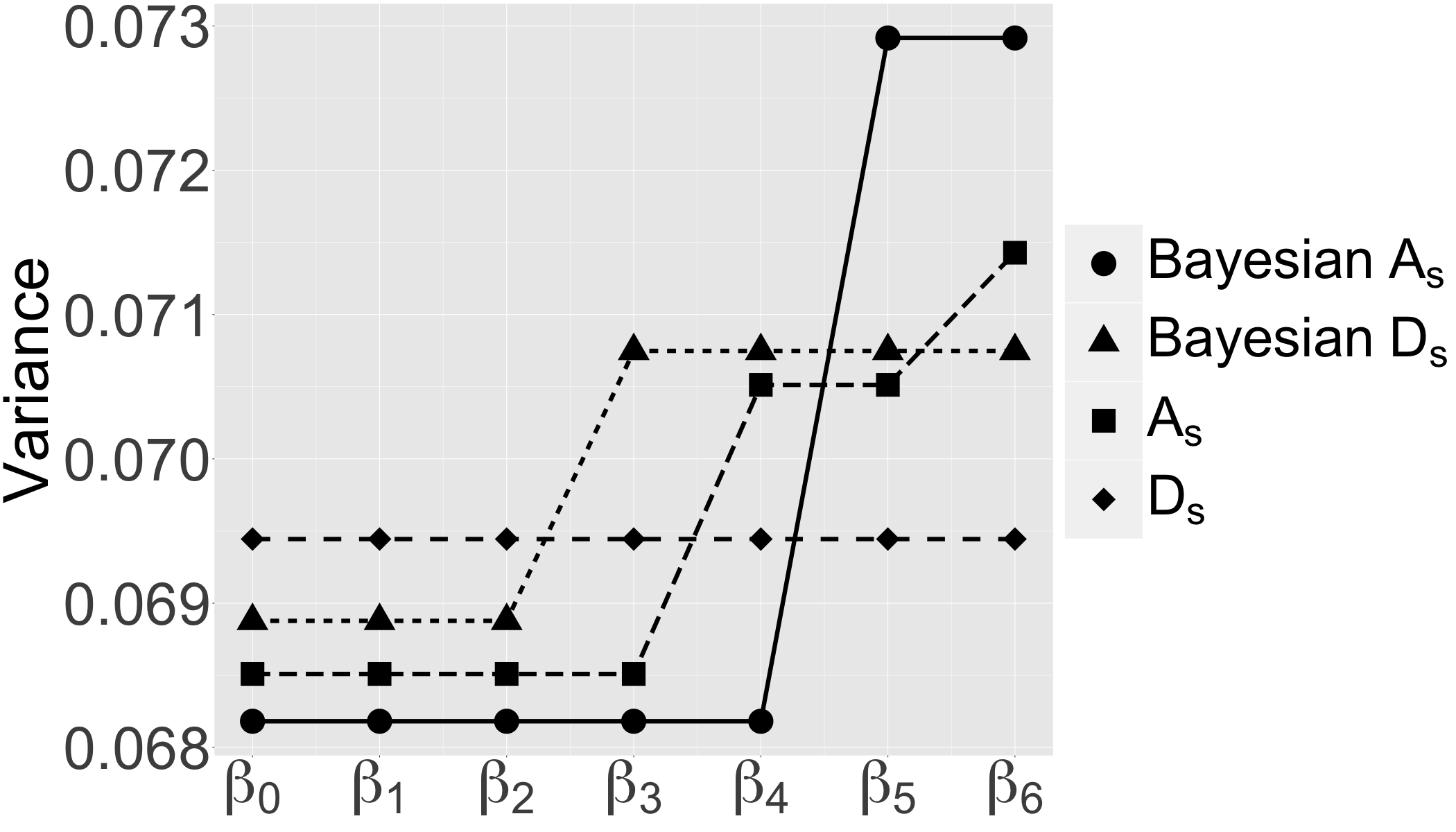}}}$
\end{minipage}
\hfill
\begin{minipage}[b]{.5\textwidth}
\centering
$\vcenter{\hbox{\includegraphics[width=1\textwidth]{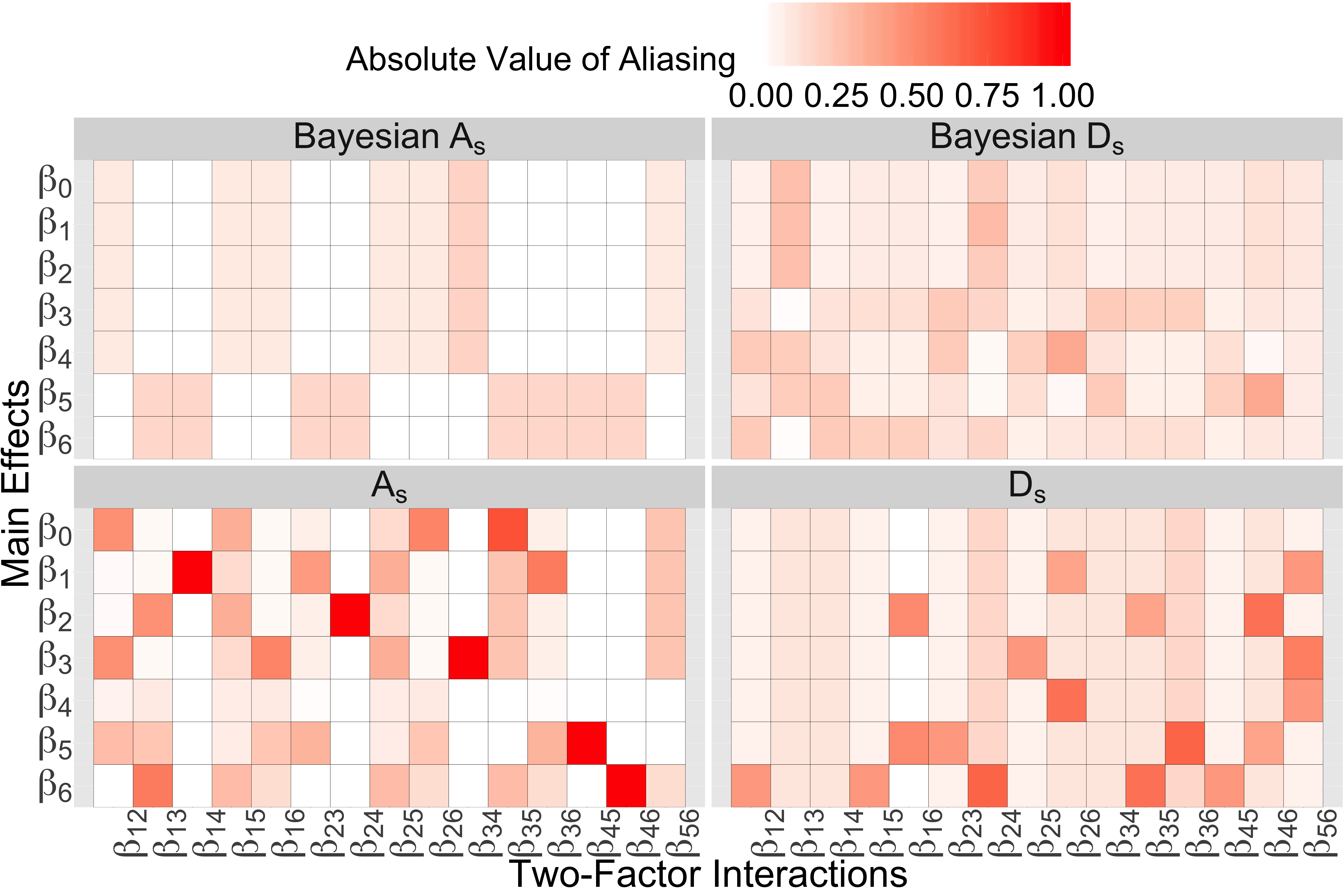}}}$
\end{minipage}
\caption{(Left) Variances under main effect model for four designs with $n=15$ and $k=6$. (Right) Heatmap of alias matrices in absolute value for main effects .\label{fig:6F15RVariances}}
\end{figure}

Figure~\ref{fig:6F15RVariances} depicts variances (in ascending order) under the main effect model and alias matrices for the Bayesian $A_s$- and Bayesian $D_s$-optimal designs, as well as the $A_s$- and $D_s$-optimal designs generated in Section~4.1.  The displayed alias matrices only show aliasing of the main effects.  The Bayesian $A_s$-optimal design was found for all $15 \leq \tau^{-2} \leq 100$ and had settings $x_{ij} \in \{\pm 1, 0\}$.  The Bayesian $A_s$-optimal design for $\tau^{-2} = 10$ had smaller aliasing (as measured by $\text{tr}(\boldsymbol{A}^T\boldsymbol{A})$, following \cite{jones2011efficient}) and had non-integer settings. The design is provided in the Supplementary Materials.  The Bayesian $D_s$-optimal design shown in Figure~\ref{fig:6F15RVariances} is comprised of $x_{ij}=\pm 1$ and was found for all $20 \leq \tau^{-2} \leq 100$.  It was chosen due to its minimizing $\text{tr}(\boldsymbol{A}^T\boldsymbol{A})$ among all constructed Bayesian $D_s$-optimal designs.  

The $D_s$-optimal design estimates all main effects with equal variance, while the $A_s$-optimal design has smaller variances except for $\hat{\beta}_6$.  The Bayesian $A_s$-optimal design has both the smallest and largest individual variances. The Bayesian $A_s$- and $D_s$-optimal designs have superior aliasing properties over their non-Bayesian counterparts. The Bayesian $A_s$-optimal design minimized $\text{tr}(\boldsymbol{A}^T\boldsymbol{A})$ compared to the other three designs. This reduced aliasing can in part be attributed to the $0$ coordinates. When $x_{ij}=\pm 1$, a design with an odd number of runs will necessarily have some degree of column correlation. A design having some $x_{ij}=0$ can achieve orthogonality between columns for such $n$ and hence zero elements in the alias matrix. Orthogonality through including $x_{ij}=0$, however, leads to larger variances for the associated main effects. 

\subsection{Screening Quadratic Models}\label{s:RSM}

Effect principles applied to a quadratic model leads to the partitioning of main effects as primary terms and potential terms of all two-factor interaction and quadratic effects.  We assigned different Bayesian priors to the two sets of potential effects, letting $\tau_I^{-2}$ and $\tau_Q^{-2}$ denote the prior precision for the two-factor interaction and quadratic effects, respectively. We constructed Bayesian $A_s$- and $D_s$-optimal designs under $\tau_I^{-2} \in \{1,16\}$ and $\tau_Q^{-2} \in \{0,1,16\}$ using $10,000$ starting designs.  For $\tau_Q^{-2}=0$, the quadratic effects become primary terms. We considered $k = 6, 8, 10$ and $n=(2k+1),\dots,(1+k+k^2)$.  The minimum $n$ value considered is that for a definitive screening design \citep{jones2011class} and the last is a run size that allows estimation of the full model.  A definitive screening design (DSD) has $k$ foldover pairs of $\boldsymbol{x}_i$, each comprised of a single zero coordinate and $k-1$ coordinates of $\pm 1$.  DSDs have no aliasing of the main effects with the interaction and quadratic terms.

For a given design, let $\boldsymbol{F}_M$, $\boldsymbol{F}_I$, and $\boldsymbol{F}_Q$ be the model matrices corresponding to the main effects, interactions, and quadratic effects, respectively. Each design is summarized using three metrics: (1) $\log(A_{M})$ where $A_M$ is the sum of the main effect variances for a fitted main effect model; (2) $\log(SS_Q)$ where $SS_Q$ is the sum of squared off-diagonals of $\boldsymbol{F}_Q^T\boldsymbol{F}_Q$, and (3)  $\log(SS_{MI}+1)$ where $SS_{MI}$ is the sum of squared values of $\boldsymbol{F}_M^T\boldsymbol{F}_I$.  The metrics $\log(SS_Q)$ and $\log(SS_{MI}+1)$ are surrogates for the information dedicated to quadratic effects and aliasing between main effects and interactions, respectively.

\begin{figure}
\centering
\includegraphics[width=0.95\textwidth]{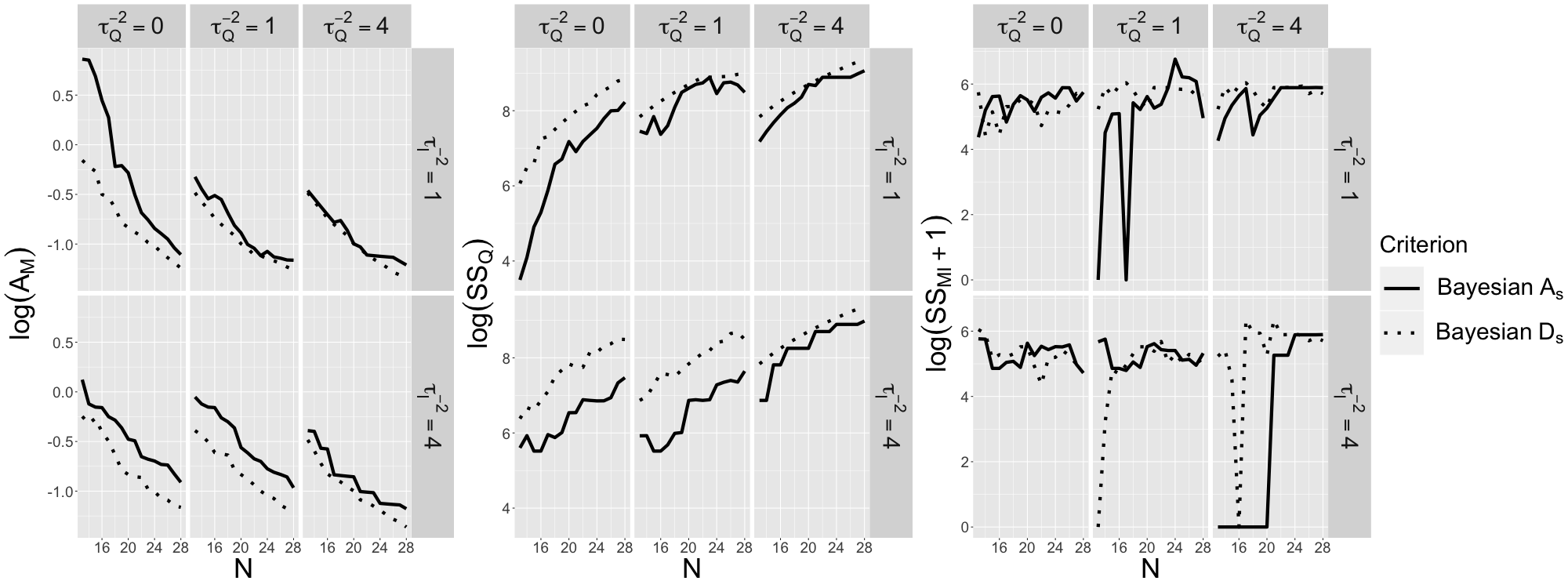}
\caption{Performance measures for best Bayesian-$D_s$ and -$A_s$ designs when $k = 6$ found with $\tau_Q^{-2} \in \{0,1,16\}$ and $\tau_I^{-2} \in \{1,16\}$. (Left) The $A_s$-criterion for the main effect model on the log scale.  (Middle) The sum of squares of the off-diagonals for the quadratic terms on the log scale. (Right) The sum of squares of the cross products of the main effects and interactions on the log scale with offset $1$.\label{fig:BayesRSMk6}}
\end{figure}

Figure~\ref{fig:BayesRSMk6} shows the numerical results for $k=6$ factors; similar conclusions were reached for the $k=8$ and $k=10$ scenarios (see Supplementary Materials).  Generally, the Bayesian $A_s$-optimal design's variances under the main effect model were worse than those under the Bayesian $D_s$-optimal design.  However, for fixed values $\tau_Q^{-2}$ and $\tau_{I}^{-2}$, the Bayesian $A_s$-optimal design had comparable or smaller values for $\log(SS_Q)$ and $\log(SS_{MI}+1)$, implying better estimation capacity and aliasing properties for the potential effects.  The Bayesian $A_s$-optimal designs for $\tau_Q^{-2}=\tau_I^{-2}=16$ closely resemble the structure of DSDs for $n=13,\dots,20$ with no aliasing between main effects and interactions.  The Bayesian $A_s$-optimal designs for $n=13$ and $17$ were a DSD and augmented DSD \citep{JonesNachtsheim2017}, respectively. For $n=14$ and $n=18,19,20$, the Bayesian $A_s$-optimal designs added center runs (i.e., $\boldsymbol{x}_i=0$) to the DSD and augmented DSD, respectively.  The Bayesian $A_s$-optimal design for $n=15$ had one center run and 7 pairs of foldover runs, mimicking the DSD structure.  The Bayesian $D_s$-optimal designs were less likely to identify designs with structures similar to DSDs for the $\tau_Q^{-2}$ and $\tau_{I}^{-2}$ we considered.

The behavior of the Bayesian $A_s$-optimal designs was influenced by the measure's implicit emphasis on quadratic effects due to their estimators having larger minimum variances than main effects and interactions.  This phenomenon was mentioned for the $A_s$-criterion by \cite{gilmour2012optimum} and discussed thoroughly by \cite{AllenMoyer2021}.  To equally emphasize minimizing variances among all effects, both articles recommend an $A_W$-criterion that incorporates the minimum variances.  \cite{AllenMoyer2021} refer to this $A_W$-criterion as the standardized $A_w$-criterion.  Note, this modification is unnecessary for main effect and interaction models because these effects have the same minimum variance.  Extending the weighted approach by \cite{AllenMoyer2021} to the Bayesian $A_s$-criterion requires the introduction of a weight matrix based on the minimum posterior variances for given prior variances.  For the quadratic model, this would lead to a diagonal weight matrix with smaller weights assigned to quadratic effects.  However,
$\tau_Q^{-2}$ also controls the magnitude of the quadratic effects' posterior variances so manipulating posterior variances for potential terms via weighting can be done equivalently through manipulation of $\tau_Q^{-2}$. Indeed, in Figure~\ref{fig:BayesRSMk6} we see that as $\tau_Q^{-2}$ increases, $\log(A_M)$ decreases and $\log(SS_Q)$ increases implying less focus on quadratic effects.  We would expect the same behavior if we were to assign smaller weights to the quadratic effects.

\section{A Blocked Screening Design for Vitamin Photosensitivity}\label{s:block}

\cite{GoosJones} discuss a blocked screening experiment performed by a pharmaceutical manufacturer that aimed to determine a combination of vitamins and certain fatty molecules to reduce the vitamins' photosensitivity, thereby increasing the product's shelf life. There were six factors studied corresponding to the presence/absence of riboflavin as well as five fatty molecules. The measuring device required daily recalibration, allowing only four measurements per day. The experiment was broken up across eight days that allowed $4$ runs per day, leading to a study of $k=6$ factors and $b=8$ blocks each of size $4$. The experimenters wanted to be able to estimate all six main effects and 15 two-factor interactions because they were concerned about possible large interactions.

Many of the techniques for constructing fractional factorial designs can be employed to create blocked screening experiments but only for certain values of $b$ and $u$.  For example, if $n=bu=2^k$ and $b=2^\ell$, we can block all $2^k$ treatment combinations by confounding $2^\ell - 1$ factorial effects with the block effects.  All remaining factorial effects are estimated with minimal variance.  If $n=bu=2^{k-m}$ and $b=2^\ell$, then we may block a fractional factorial design based on certain confounding patterns \citep{Bisgaard1994,chen1999,cheng2001,Cheng2002}.  However, \cite{Cheng2004} demonstrate nonregular fractional factorial design may have superior estimation and variance properties.  For the vitamin photosensitivity experiment, a blocked regular fractional factorial will not be able to estimate all two factor interactions, an important property for their application.

\begin{figure}[ht]
\begin{minipage}[b]{.49\textwidth}
\centering
\scalebox{0.6}{\begin{tabular}{rrrrrrrrrrrrr}
-1 & -1 & -1 & 1 & 1 & -1 & & 1 & -1 & -1 & 1 & -1 & -1 \\ 
  1 & 1 & 1 & 1 & 1 & 1 & & 1 & 1 & 1 & -1 & 1 & -1 \\ 
  -1 & 1 & 1 & -1 & -1 & -1 & & -1 & 1 & -1 & -1 & -1 & 1 \\ 
  1 & -1 & -1 & -1 & -1 & 1 & & -1 & -1 & 1 & 1 & 1 & 1 \\ \\
  1 & -1 & 1 & 1 & -1 & 1 & & 1 & 1 & -1 & 1 & -1 & 1 \\ 
  -1 & -1 & -1 & -1 & 1 & 1 & & -1 & -1 & 1 & 1 & -1 & -1 \\ 
  1 & 1 & -1 & -1 & -1 & -1  & & -1 & 1 & -1 & -1 & 1 & -1 \\ 
  -1 & 1 & 1 & 1 & 1 & -1 & & 1 & -1 & 1 & -1 & 1 & 1 \\ \\
  -1 & 1 & -1 & 1 & -1 & -1 & & -1 & -1 & -1 & -1 & -1 & -1 \\ 
  1 & -1 & 1 & -1 & -1 & -1 & & 1 & 1 & -1 & -1 & 1 & 1 \\ 
  -1 & 1 & 1 & -1 & 1 & 1 & & 1 & -1 & 1 & 1 & 1 & -1 \\ 
  1 & -1 & -1 & 1 & 1 & 1 & & -1 & 1 & 1 & 1 & -1 & 1 \\ \\
  -1 & 1 & -1 & 1 & 1 & 1 & & 1 & 1 & -1 & 1 & 1 & -1 \\ 
  1 & -1 & -1 & -1 & 1 & -1 & & 1 & 1 & 1 & -1 & -1 & 1 \\ 
  1 & 1 & 1 & 1 & -1 & -1  & & -1 & -1 & -1 & 1 & -1 & 1 \\ 
  -1 & -1 & 1 & -1 & -1 & 1 & & -1 & -1 & 1 & -1 & 1 & -1 \\ 
\end{tabular}}
\end{minipage}
\begin{minipage}[b]{.4\textwidth}
\centering
$\vcenter{\hbox{\includegraphics[width=1\textwidth]{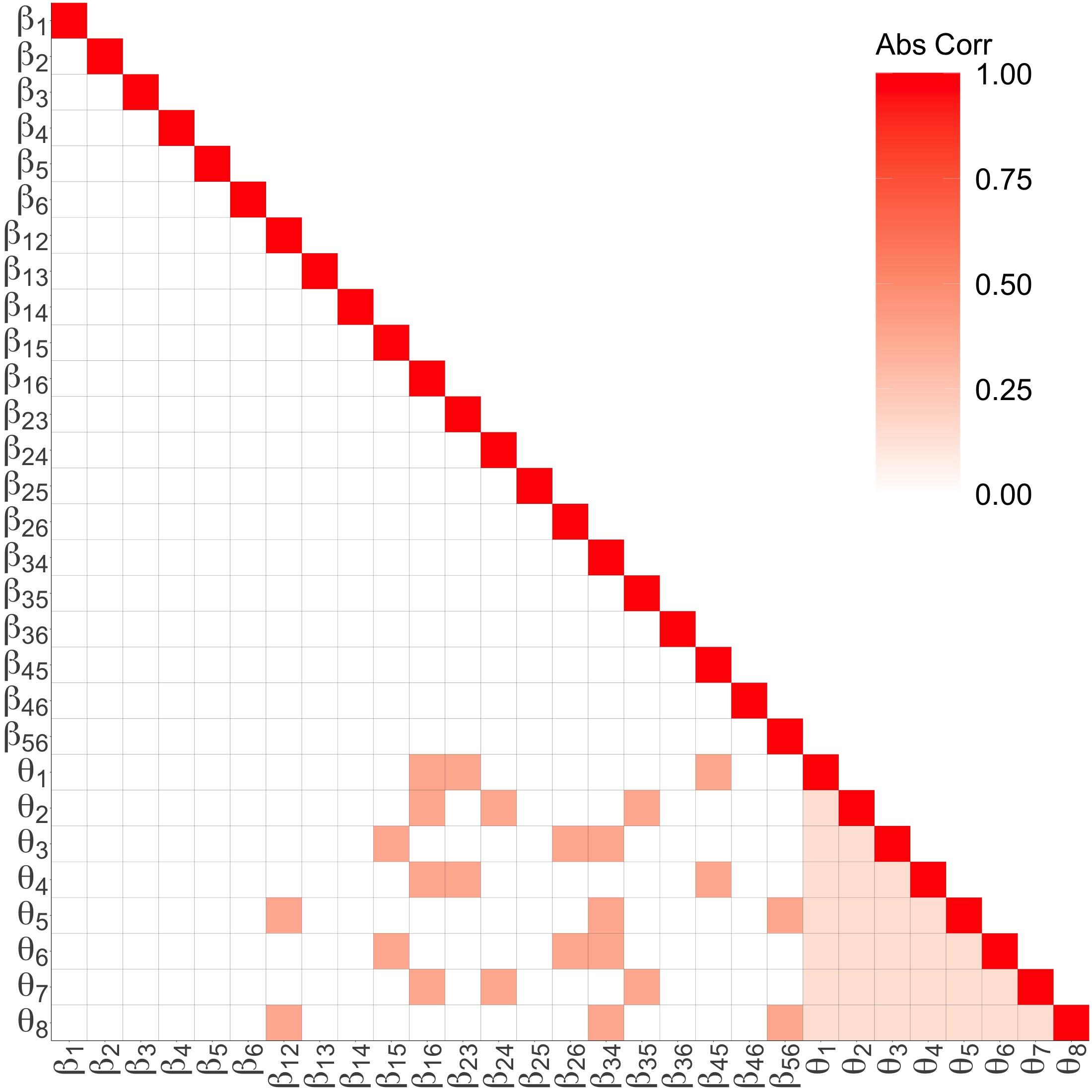}}}$
\end{minipage}
\caption{(Left) Blocked screening design with 6 factors within 8 blocks of size 4 and (Right) heatmap of $L^TL$.  Each group of the block design represents one block with $4$ runs and settings for $6$ factors.\label{fig:BlockDesign}}
\end{figure}

 \cite{GoosJones} constructed a $D$-optimal blocked design algorithmically that can estimate all main effects and two-factor interaction effects.  The experiment had only categorical factors, but we will treat them here as if they were continuous. We constructed blocked designs with the $D_s$-criterion, $A_s$-criterion, and a Bayesian $A_s$-criterion with $\tau_I^{-2}=16$.  The block effects were assigned weight $w=10^{-6}$ and $10,000$ starting designs were used.  Although 3 different designs were constructed corresponding to the different criteria, the best design found under the Bayesian $A_s$-criterion, shown in Figure~\ref{fig:BlockDesign}, was optimal across all criteria.  Even after increasing the number of starting designs to $100,000$, the CEAs for the $D_s$- and $A_s$-criteria were still unable to identify this design.

The optimal design we constructed is more $D$-efficient than the design reported in \cite{GoosJones}, and consists entirely of $\pm 1$ coordinates and so can be used in their application.  The design turned out to be a nonregular fractional factorial with generalized resolution $3.75$ and has only 6 words of length $4.5$. \cite{Cheng2004} tabulated designs with similar structure but only for $n=12, 16$, and $20$.  Others have looked at minimizing generalized aberration for larger run sizes (see \cite{FANG2007740} and \cite{Schoen2017}), but not in the context of blocking.  The main effects of the design are all estimated with optimal variance ($1/32$) and have zero aliasing with the block and interaction effects.  Five of the 15 interactions are also estimated with optimal variance.  The remaining interaction effects are partially correlated with the block effects, with 8 of the interactions having a variance of $0.047$ and the other two having a variance of $0.063$.

\section{Discussion}\label{s:Discussion}

This paper compares different forms of the $D$- and $A$-criterion for constructing screening designs that simultaneously minimize variances and bias of a predetermined set of effects in a linear model.  We challenge two commonly held beliefs concerning screening designs: 
\begin{itemize}
    \item Algorithmic optimization of the $D$-criterion produces good screening designs for arbitrary $n$.
    \item When constructing screening designs, one needs only to consider $x_{ij}=\pm 1$ even if $x_{ij}$ is numeric and can take on other values in $[-1,1]$.
\end{itemize} 
\cite{gilmour2012optimum} and \cite{jones2020Aoptimal} have also pointed out the failing of the $D_s$-criterion, and we have further clarified these failings.  Our investigation of the $D_s$-criterion's CEA shows that many $D_s$-optimal designs can exist for a given scenario, having different variance and bias properties.  The superior performance of our continuous CEA in Section~\ref{s:CoordExchange} indicates that even if an $A_s$-optimal design is comprised of only $x_{ij} \in \{\pm 1, 0\}$, the continuous CEA more frequently constructs an $A_s$-optimal design than a discrete CEA.  Moreover, we also found some combinations of $n$ and $k$ where the $A_s$-optimal design included non-integer coordinates.

Our investigation of Bayesian $A_s$- and Bayesian $D_s$-optimal designs in Sections~\ref{subsec:2fibayes} and \ref{s:RSM} revealed that the Bayesian $A_s$-criterion better balances variance and bias minimization for the prior variance values considered.  In Section~5, we found that the optimal design constructed under the Bayesian $A_s$-criterion was also optimal under the $A_s$- and $D_s$-criterion, and that it was better than the designs constructed directly under these two criteria.  This is unfortunately a possibility with algorithmic construction and we recommend  practitioners generate multiple designs under different design criteria and compare them by inspecting their variances and biases directly, similar to \cite{AllenMoyer2021}.  

There are many directions of future research we are currently investigating.  First, this paper is combines weighted and Bayesian criteria, but mainly to ignore the variances of nuisance effects.  Section~\ref{s:RSM} hinted at redundancies in weighting posterior effects but there may be opportunities for more flexible weighting applied to primary effects.  Next, more investigation is needed to compare the Bayesian $A_s$-criterion to more brute force methods that minimize variance and bias. These methods commonly employ some type of $D$-criterion to measure variance which could easily be modified to be an $A$-criterion. Following \cite{Li2006}, more investigation is needed on the difference of the optimal designs under the $\mathcal{D}$- and $\mathcal{A}$-criteria when higher-order interactions are considered. Finally, following \cite{gilmour2012optimum} and \cite{JonesGOSSD}, we are currently developing an $A_s$-criterion that includes external variance estimation through replication or fake factors.


\section{Supplementary Materials}

\subsection{Theorem 1 Proof}\label{A:Thm1prof}
Under an $m$-factor interaction model, the elements of $f_1(x_{ij})$ are comprised of $x_{ij}$ and products of $x_{ij}$ with the other main effect coordinates.  Hence $f_1(\tilde{x}) = \tilde{x} \times f_{(1)}(\boldsymbol{x}_{i,-j})$ where $f_{(1)}(\boldsymbol{x}_{i,-j})$ is a vector whose elements depend only on $\boldsymbol{x}_{i,-j}$.  For simplicity, we write $f_{(1)}$ instead of $f_{(1)}(\boldsymbol{x}_{i,-j})$. The definition coordinate update formula, $\Delta_D^{ij}$, for the $D$- and $D_s$-criterion involves matrix $\boldsymbol{D}=(\boldsymbol{L}^T\boldsymbol{L})^{-1}$, and for the Bayesian versions of these criteria $\boldsymbol{D}=(\boldsymbol{L}^T\boldsymbol{L}+\tau^{-2}\boldsymbol{K})^{-1}$. Recall $\boldsymbol{V}=(1-v(\boldsymbol{x}_i))\boldsymbol{D}+\boldsymbol{D}l(\boldsymbol{x}_i)l^T(\boldsymbol{x}_i)\boldsymbol{D}$ where $v(\boldsymbol{x}_i)=l^T(\boldsymbol{x}_i)\boldsymbol{D}\,l(\boldsymbol{x}_i)$. Partition $\boldsymbol{V}$ so $\boldsymbol{V}_{11}$ corresponds to elements of $f_1(\tilde{x})$ and $\boldsymbol{V}_{22}$ corresponds to $l_2(\boldsymbol{x}_{i,-j})=(f_2(\boldsymbol{x}_{i,-j})^T,\boldsymbol{z}_i^T)^T$. Then
\[
\Delta^{ij}_D(\tilde{x}) = \tilde{x}^2f^T_{(1)}\boldsymbol{V}_{11} f_{(1)} + \tilde{x}\boldsymbol{a}^Tf_{(1)} + c
\]
is a quadratic polynomial where $\boldsymbol{a}=2\boldsymbol{V}_{12}l_2(\boldsymbol{x}_{i,-j})$ and $c =l^T_2(\boldsymbol{x}_{i, -j})\boldsymbol{V}_{22}l_2(\boldsymbol{x}_{i, -j}) + \{1 - v(\boldsymbol{x}_i)\}$. The submatrix $\boldsymbol{V}_{11}$ is positive semidefinite so $f^T_{(1)}\boldsymbol{V}_{11} f_{(1)} \geq 0$, making $\Delta^{ij}_D(\tilde{x})$ a convex quadratic polynomial with respect to $\tilde{x}$.  Hence either $\tilde{x}=-1$ or $1$ maximizes $\Delta^{ij}_D(\tilde{x})$ across $\tilde{x} \in [-1,1]$

\subsection{Corollary 1 Proof}
For the scenario in Theorem 1, if an optimal design includes an $x_{ij} \neq \pm 1$, then Theorem 1 tells us that there exists another $D$-optimal design by exchanging this $x_{ij}$ with either $-1$ or $1$. Denote this equivalent exchange by $\tilde{x}^*$. Theorem~1 tells us that $\Delta_D^{ij}$ is a convex quadratic polynomial and, since $\Delta_D^{ij}(x_{ij})=\Delta_D^{ij}(\tilde{x}^*)=1$ the following properties must hold:
\begin{enumerate}
    \item For all $\tilde{x}$ between $x_{ij}$ and $\tilde{x}^*$, $\Delta_D^{ij}(\tilde{x})\leq 1$
    \item For all $\tilde{x}$ not between $x_{ij}$ and $\tilde{x}^*$, $\Delta_D^{ij}(\tilde{x})\geq 1$\ .\
\end{enumerate}
Since $x_{ij} \neq \pm 1$, case (2) must have $\Delta_D^{ij}(\tilde{x})= 1$ for all such $\tilde{x}$, otherwise we could find a $\tilde{x}$ that would improve over the initial optimal design, a contradiction. Since $\Delta_D^{ij}(\tilde{x})= 1$ for all $\tilde{x}$ in this nonempty interval, $\Delta_D^{ij}$ must be constant across all $[-1,1]$, meaning all possible exchanges will produce $D$-optimal designs.

\subsection{$A_W$-criterion Coordinate Exchange Formula}\label{A:CEA}

We first derive the row exchange formulas for the weighted $A$-criterion $\text{tr}\{\boldsymbol{W}(\boldsymbol{L}^T\boldsymbol{L})^{-1}\}$ for a given positive definite matrix $\boldsymbol{W}$.  Note $\boldsymbol{W}=\boldsymbol{I}$ yields the traditional $A$-criterion.  Define
\begin{align}
\Delta_{AW}(\boldsymbol{x}_i, \tilde{\boldsymbol{x}})=\text{tr}\{\boldsymbol{W}(\boldsymbol{L}_0^T\boldsymbol{L}_0)^{-1}\} - \text{tr}\{\boldsymbol{W}(\widetilde{\boldsymbol{L}}^T\widetilde{\boldsymbol{L}})^{-1}\}\ .\
\end{align}
Since $\widetilde{\boldsymbol{L}}^T\widetilde{\boldsymbol{L}}=\boldsymbol{L}_0^T\boldsymbol{L}_0 + \boldsymbol{L}_{01}\boldsymbol{L}_{02}^T$ for $\boldsymbol{L}_{01} = (l(\tilde{\boldsymbol{x}}), -l(\boldsymbol{x}_i))$ and $\boldsymbol{L}_{02} = (l(\tilde{\boldsymbol{x}}), l(\boldsymbol{x}_i))$, it follows by \cite{sherman1950adjustment}
\begin{align}
(\widetilde{\boldsymbol{L}}^T\widetilde{\boldsymbol{L}})^{-1}=\boldsymbol{D} - \boldsymbol{D}\boldsymbol{L}_{01}(\boldsymbol{I}+\boldsymbol{L}_{02}^T\boldsymbol{D}\boldsymbol{L}_{01})^{-1}\boldsymbol{L}_{02}^T\boldsymbol{D}\ ,\
\end{align}
where $\boldsymbol{D}=(\boldsymbol{L}_{0}^T\boldsymbol{L}_{0})^{-1}$.  With $\phi_W(\boldsymbol{x}_i)= l^T(\boldsymbol{x}_i)\boldsymbol{D}\boldsymbol{W}\boldsymbol{D}l(\boldsymbol{x}_i)$ we arrive at the expression
\begin{align*}
    \Delta_{AW}(\boldsymbol{x}_i, \tilde{\boldsymbol{x}})
   &= \text{tr}\left\{\boldsymbol{W}\boldsymbol{D}\boldsymbol{L}_{01}\left(\boldsymbol{I} + \boldsymbol{L}^T_{02}\boldsymbol{D}\boldsymbol{L}_{01}\right)^{-1}\boldsymbol{L}^T_{02}\boldsymbol{D}\right\} \\
&= \text{tr}\left\{\boldsymbol{W}\boldsymbol{D}\boldsymbol{L}_{01}\begin{pmatrix}
 1 + v(\tilde{\boldsymbol{x}})& -v(\boldsymbol{x}_i, \tilde{\boldsymbol{x}})  \\
 v(\boldsymbol{x}_i, \tilde{\boldsymbol{x}}) & 1 - v(\boldsymbol{x}_{i})
\end{pmatrix}^{-1}\boldsymbol{L}^T_{02}\boldsymbol{D}\right\} \\
&= \frac{1}{\Delta_D(\boldsymbol{x}_i, \tilde{\boldsymbol{x}})} \text{tr}\left\{\boldsymbol{W}\boldsymbol{D}\boldsymbol{L}_{01}\begin{pmatrix}
 1 - v(\boldsymbol{x}_{i})& v(\boldsymbol{x}_i, \tilde{\boldsymbol{x}})  \\
 -v(\boldsymbol{x}_i, \tilde{\boldsymbol{x}}) & 1 + v(\tilde{\boldsymbol{x}})
\end{pmatrix} \boldsymbol{L}^T_{02}\boldsymbol{D}\right\} \\
&=  \frac{1}{\Delta_D(\boldsymbol{x}_i, \tilde{\boldsymbol{x}})} \{l^T(\tilde{\boldsymbol{x}})\boldsymbol{U}l(\tilde{\boldsymbol{x}}) - \phi_W(\boldsymbol{x}_i) \} \ ,\
\end{align*}
where $v(\boldsymbol{x}_i,\tilde{\boldsymbol{x}})=\boldsymbol{x}_i^T\boldsymbol{D}\tilde{\boldsymbol{x}}$ and $$\boldsymbol{U} = \{1 - v(\boldsymbol{x}_i)\}\boldsymbol{D}\boldsymbol{W}\boldsymbol{D} + \boldsymbol{D} l(\boldsymbol{x}_i) l^T(\boldsymbol{x}_i)\boldsymbol{D}\boldsymbol{W}\boldsymbol{D} +\boldsymbol{D} \boldsymbol{W}\boldsymbol{D} l(\boldsymbol{x}_i) l^T(\boldsymbol{x}_i)\boldsymbol{D} - \phi_W(\boldsymbol{x}_i)\boldsymbol{D}\ .\ $$ The coordinate exchange formulas follow by straightforward partitioning and permuting.

\subsection{CEA of $A_W$-criterion for $m$-factor interaction model}\label{A:Alvl}
For $x_{ij} \in [-1,1]$, let $q \in \mathbb{R}$ and define
\begin{align*}
    F(q) &= \max_{\tilde{x} \in [-1,1]}\left\{l^T_1(\tilde{x})(\boldsymbol{U}_{11}-q\boldsymbol{V}_{11}) l_1(\tilde{x}) + (\boldsymbol{b}-q\boldsymbol{a})^T l_1(\tilde{x}) + (d - qc)\right\}\\
    a(q) &= l^T_{(1)}(\boldsymbol{U}_{11} - q\boldsymbol{V}_{11})l_{(1)}\\
    b(q) &= (\boldsymbol{b}^T - q\boldsymbol{a}^T)l_{(1)}\\
    c(q) &= d - qc\ .\
\end{align*} 
For $q_0 = \underset{\tilde{x} \in [-1, 1]}{\max} \Delta^{ij}_{AW}( \tilde{x})$, define
\begin{equation*}
G(\tilde{x}) = \tilde{x}^2a(q_0) + \tilde{x}b(q_0) + c(q_0).
\end{equation*}
Applying the main theorem of \cite{dinkelbach1967nonlinear}, we note that $G(\tilde{x}) \leq 0$ for all $\tilde{x} \in [-1, 1]$ and $\tilde{x}^*$ optimizes $\Delta^{ij}_A$ if and only if $G(\tilde{x}^*) = 0$. To identify the optima of $\Delta^{ij}_{AW}$, it is sufficient to identify the optima of $G(\tilde{x})$, a quadratic function whose optima clearly depend on the sign of $a(q_0)$ and $b(q_0)$.  We enumerate the possible optima here:
\begin{enumerate}
    \item $a(q_0) > 0$, $\tilde{x}^* = -1$ or $1$
    \item $a(q_0) = 0$ 
    \begin{enumerate}
        \item If $b(q_0) \neq 0$, $\tilde{x}^* = -1$ or $1$
        \item If $b(q_0) = 0$, any value $\tilde{x}^*$ optimizes $\Delta^{ij}_{AW}$ 
    \end{enumerate}
    \item $a(q_0) < 0$
       \begin{enumerate}
        \item If $b(q_0) \neq 0$, $\Delta^{ij}_A$ maximized at $\tilde{x}^* = \frac{-b(q_0)}{2a(q_0)},$ when $|\frac{-b(q_0)}{2a(q_0)}| \leq 1$, otherwise it is maximized at either $\tilde{x}^*=-1$ or $1$.
        \item If $b(q_0) = 0$, $\tilde{x}^* = 0$.
    \end{enumerate}
\end{enumerate}
An example of a non-integer optima is shown next.

\subsection{Demonstration of continuous optimization of  $\Delta^{ij}_{AW}$}

\begin{figure}[ht]
\begin{minipage}[b]{.45\textwidth}
\centering
    \begin{tabular}{rrrr}
        -0.10 & -1 &\fbox{-0.43}& -1 \\ 
         1  & 1 & 0.53 & -1 \\ 
        -1 & -0.25 & 1 & 0.98 \\ 
        1 & -1 & -0.43 & 1 \\ 
        1 & 1 & 1 & 1 \\ 
        1 & 1 & -1 & 1 \\ 
       -1 & 1 & -1 & 1 \\ 
       -1 & 1 & -1 & -1 \\
    \end{tabular}
\end{minipage}
\begin{minipage}[b]{.45\textwidth}
\centering
$\vcenter{\hbox{\includegraphics[width=0.95\textwidth, angle = 270]{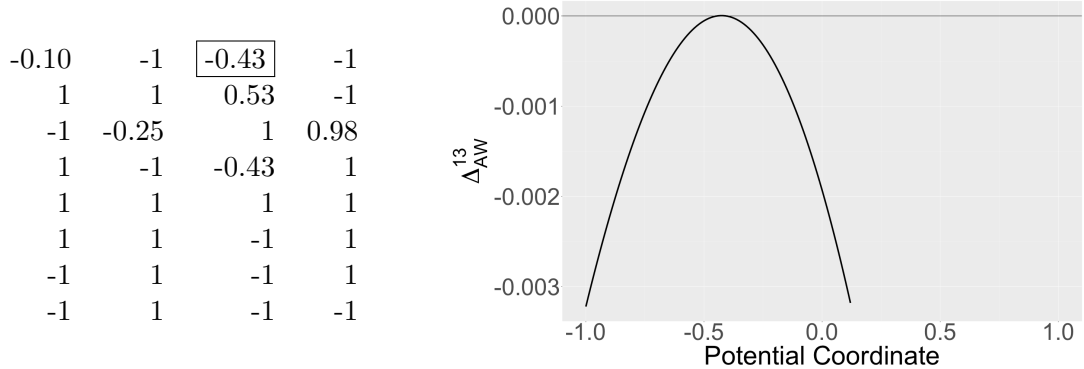}}}$
\end{minipage}
\caption{(Left) For a $n = 8,\ k = 4$ design, the objective function (right) $\Delta_{AW}^{13}$ for any replacement of the coordinate in row $1$, column $3$ indicates an exchange of $-0.45$ with the almost equivalent value of $-0.43$ optimizes $\Delta_{AW}^{13}$. \label{fig:DeltaLNonIntSwap}}
\end{figure}

We consider an $n = 8, \ k = 4$ design for a main effect model. We know that the final $8$ run, $4$ factor design can be constructed by selecting 4 factor columns from an $8 \times 8$ Hadamard matrix. However, during design construction, non-integer exchanges may be optimal. For example, in the middle of a continuous CEA where the coordinate $0.45$ in the $1$st row and $3$rd column of the design in Figure~\ref{fig:DeltaLNonIntSwap} is being exchange with a new optima, by computing $a(q)$, $b(q)$, and $c(q)$ for the original design, we get 

\begin{align*}
    a(q) &= -0.0104 - q0.07  \\
    b(q) &= -0.0089 + q0.02 \\
    c(q) &= -0.002 - q0.98 \ .\
\end{align*} 

\noindent Using \cite{dinkelbach1967nonlinear}, let $q_0 = \underset{\tilde{x} \in [-1, 1]}{\max} \Delta^{13}_A(x_{ij}, \tilde{x}, \boldsymbol{x}_{i, -j})$. Start with a guess for $q_0$ and $\tilde{x}$ and compute $F(q_0)$ and $\Delta_A^{13}$ iteratively until they converse. In this example, the true $q_0 \approx 0$, so $a(q) < 0$, $b(q) < 0$, meaning the optimum value is at $\frac{-b(q)}{2a(q)} \approx \frac{-0.0089}{2 \times 0.0104} \approx -0.43$.

\begin{figure}[h]
\begin{minipage}[b]{.45\textwidth}
\centering
    \begin{tabular}{rrrr}
          -1 & 1 & -1 & -1\\
          0.44 & 1 & 1 & 1\\
          -0.64 & -1 & 1 & -1\\
         \rowcolor{lightgray}  1 & -1 & -1 & 0.17\\
          1 & 1 & -1 & 0\\ 
           -1 & 0 & -1 & 1\\
    \end{tabular}
\end{minipage}
\begin{minipage}[b]{.45\textwidth}
\centering
$\vcenter{\hbox{\includegraphics[width=.8\textwidth, angle = 270]{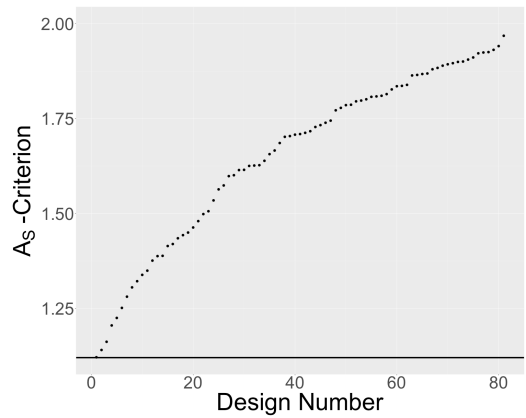}}}$
\end{minipage}
 \caption{Row 4 of the design (left) exchanges $(1, -1, -1, 0)$ with $(1, -1, -1, 0.17)$. (Right) A-values of every integer-only row exchange compared with the value at the optimal exchange, represented by the horizontal line. Only the original row has an A-value close to the value of the optimal switch ($1.121$ verses $1.122$).\label{tab:RowExcEx}}
\end{figure}

The same approach holds for a row exchange. Figure~\ref{tab:RowExcEx} shows a $4$ factor $6$ run design with non-integer coordinates.  This design is $\widetilde{\boldsymbol{X}}_d$ after exchanging the original row $\boldsymbol{x}_4=(1,\ -1, \ -1, \ 0)$ with the row $(1, -1, -1, 0.17)$.  None of the $3^4$ integer-only row exchanges yielded an improved value of the objective function. 

\clearpage

\subsection{Proof of Corollary 2}
It easy to show that, in general, $\boldsymbol{V}\, l(\boldsymbol{x}_i)=\boldsymbol{D} \, l(\boldsymbol{x}_i)$ and
\[
\boldsymbol{V}\boldsymbol{W}\boldsymbol{D}=(1-v(\boldsymbol{x}_i))\boldsymbol{D}\boldsymbol{W}\boldsymbol{D}+\boldsymbol{V}l(\boldsymbol{x}_i) l^T(\boldsymbol{x}_i)\boldsymbol{V}\boldsymbol{W}\boldsymbol{D}\ .\
\]
Then we have the following simplified expression for $\boldsymbol{U}$:
\[
\boldsymbol{U} = \boldsymbol{V}\boldsymbol{W}\boldsymbol{D} + \boldsymbol{D}\boldsymbol{W}\boldsymbol{V} \,  l(\boldsymbol{x}_i)\, l^T(\boldsymbol{x}_i)
\boldsymbol{V}-\phi_W(\boldsymbol{x}_i) \boldsymbol{D}\ .\
\]
For the situation described in Corollary 1, the coordinate update for each criterion among the $\mathcal{A}$-criteria involves some $\Delta_D^{ij}$ in the denominator.
If this $\Delta_D^{ij}$ is constant, $f^T_{(1)}\boldsymbol{V}_{11} f_{(1)}=0$ and since $\boldsymbol{V}$ is positive semidefinite, $f_{(1)}$ must be a null eigenvector for $\boldsymbol{V}_{11}$. Moreover, $f_{(1)}^T\boldsymbol{V}_{12}=f_{(1)}^T\boldsymbol{V}_{11}\boldsymbol{V}_{11}^-\boldsymbol{V}_{12}=0$. Then
\[
\boldsymbol{V}\, l(\tilde{\boldsymbol{x}})=\boldsymbol{V}\begin{pmatrix} \tilde{x} f_{(1)}\\ l_2(\boldsymbol{x}_{i,-j}) \end{pmatrix}=\begin{pmatrix}\boldsymbol{V}_{12}\, l_2(\boldsymbol{x}_{i,-j})\\\boldsymbol{V}_{22}\, l_2(\boldsymbol{x}_{i,-j}) \end{pmatrix}\ ,\
\]
which does not involve $\tilde{x}$. This and the expression for $\boldsymbol{U}$ says the coefficient for $\tilde{x}^2$ in $\Delta_{AW}^{ij}$ equals $-\phi_W(\boldsymbol{x}_i)f_{(1)}^T\boldsymbol{D}_{11} f_{(1)}$ where $\boldsymbol{D}_{11}$ is the relevant partition of $\boldsymbol{D}$, a positive definite matrix. The coefficient equals 0 if and only if either $\phi_W(\boldsymbol{x}_i)=0$ or $f_{(1)}$ is the all-zero vector. But neither can occur because the main effect coordinate is always included in the model so $f_{(1)} \neq 0$ and $\boldsymbol{D}\boldsymbol{W}\boldsymbol{D}$ is positive definite for $w>0$. Hence the coefficient must be negative and so $\Delta^{ij}_{AW}$ is a concave quadratic polynomial, having one unique maximum in $[-1,1]$. 

When adjusting for nuisance effects, we want to consider $\Delta_{A_s}^{ij}=\lim_{w \to 0} \Delta_{AW}^{ij}$. Hence if $\lim_{w\to 0} \phi_W(\boldsymbol{x}_i)>0$ the quadratic coefficient will again be negative, since $f_{(1)}^T\boldsymbol{D}_{11} f_{(1)}$ does not depend on $w$. Now $\phi_W(\boldsymbol{x}_i)=l^T(\boldsymbol{x}_i)\boldsymbol{D}\boldsymbol{W}\boldsymbol{D}l(\boldsymbol{x}_i)$ is a quadratic form of a symmetric matrix so $\lim_{w \to 0} \phi_W(\boldsymbol{x}_i)=0$ if and only if $\lim_{w\to0}\boldsymbol{W}^{1/2}\boldsymbol{D}l(\boldsymbol{x}_i)=0$. Partitioning $\boldsymbol{D}$ according to matrices $\boldsymbol{F}_0$ and $\boldsymbol{Z}$ gives the expression
\begin{align*}
    \lim_{w\to0}\boldsymbol{W}^{1/2}\boldsymbol{D}l(\boldsymbol{x}_i) = \lim_{w\to0} \begin{pmatrix}
    \boldsymbol{D}_F f(\boldsymbol{x}_i) + \boldsymbol{D}_{FZ} \boldsymbol{z}_i\\
    \sqrt{w} \boldsymbol{D}_{ZF} f(\boldsymbol{x}_i) + \sqrt{w}\boldsymbol{D}_{Z} \boldsymbol{z}_i\\
    \end{pmatrix}=
    \begin{pmatrix}
    \boldsymbol{D}_F f(\boldsymbol{x}_i) + \boldsymbol{D}_{FZ} \boldsymbol{z}_i\\
    0\\
    \end{pmatrix}\ .\
\end{align*}
So $\lim_{w \to 0} \phi_W(\boldsymbol{x}_i)=0$ if and only if $\boldsymbol{D}_F f(\boldsymbol{x}_i) + \boldsymbol{D}_{FZ} \boldsymbol{z}_i=0$, or
\[
f(\boldsymbol{x}_i)=-\boldsymbol{D}_F^{-1}\boldsymbol{D}_{FZ} \boldsymbol{z}_i\ .\
\]
For both the $A_W$- and Bayesian $A_W$-criterion, $-\boldsymbol{D}_F^{-1}\boldsymbol{D}_{FZ}\boldsymbol{z}_i=\boldsymbol{F}_0^T\boldsymbol{Z}(\boldsymbol{Z}^T\boldsymbol{Z})^{-1}\boldsymbol{z}_i$, which implies
\begin{align*}
    f(\boldsymbol{x}_i)&=\sum_{i'=1}^n f(\boldsymbol{x}_{i'})p_{z,i'i}\\
    &=\frac{1}{1-p_{z,ii}}\sum_{i' \neq i} f(\boldsymbol{x}_{i'})p_{z,i'i}\ ,\
\end{align*}
where $p_{z,i'i}=\boldsymbol{z}_{i'}^T(\boldsymbol{Z}^T\boldsymbol{Z})^{-1}\boldsymbol{z}_{i}$ are elements of the $i$-th column of $\boldsymbol{P}_Z$. For $p_{z,ii}=\boldsymbol{z}_{i}^T(\boldsymbol{Z}^T\boldsymbol{Z})^{-1}\boldsymbol{z}_{i}<1$, it also holds that 
\[
\boldsymbol{z}_i =\frac{1}{1-p_{z,ii}}\sum_{i' \neq i} \boldsymbol{z}_{i'}p_{z,i'i}\ .\
\]
Finally, this implies $l(\boldsymbol{x}_i)$ can be written as a linear combination of the other $n-1$ rows of $\boldsymbol{L}_0$ with coefficients $p_{z,i'i}/(1-p_{z,ii})$. But a constant $\Delta_D^{ij}$ implies $v(\boldsymbol{x}_i)=1$ meaning $l(\boldsymbol{x}_i)$ cannot be written as a linear combination of the other $n-1$ rows. Therefore $\lim_{w \to 0} \phi_W(\boldsymbol{x}_i)>0$ and $\Delta_{A_s}^{ij}$ is a concave quadratic polynomial.

\subsection{Section 4.2 Designs}

The $A_s$- and $D_s$-optimal designs for a main effect model with $n = 15, \ k = 6$ are shown in Table~\ref{tab:6F15R}. They do not account for the potential two-factor interaction effects, and thus lead to worse aliasing than the Bayesian $A_s$- and Bayesian $D_s$-optimal designs found in Table~\ref{tab:Bayes6F15R}. The Bayesian $A_s$-optimal design was found that better minimizes aliasing (as measured by $\text{tr}(\boldsymbol{A}^T\boldsymbol{A})$) over the Bayesian $A_s$-optimal in Table~\ref{tab:Bayes6F15R}. 

\begin{table}[h]
 \centering
 \caption{$n = 15, \ k = 6$ $A_s$- and $D_s$-optimal designs for main effect models.\label{tab:6F15R}}
\begin{tabular}{rrrrrr rr rrrrrr}
\\
\multicolumn{6}{c}{$A_s$-optimal} & & & \multicolumn{6}{c}{$D_s$-optimal} \\

-1 & 1 & 1 & -1 & 1 & 1  && & 1 & -1 & 1 & 1 & -1 & -1\\ 
  1 & 1 & -1 & -1 & -1 &  1   && & 1 & 1 & 1 & -1 & -1 & 1 \\ 
1 & -1 & -1 & 1 & 1 &  1   & && 1 & -1 & -1 & -1 & -1 & -1 \\
  1 & 1 & -1 & -1 & 1 & -1 & && -1 & -1 & 1 & 1 & -1 & 1 \\ 
 -1 & -1 & 1 & -1 & 1 &  -1   & && -1 & 1 & -1 & 1 & -1 & -1 \\
  -1 & -1 & -1 & 1 & 1 & -1  & && 1 & -1 & -1 & 1 & 1 & 1 \\ 
 1 & -1 & 1 & 1 & -1 & -1  & && -1 & 1 & 1 & 1 & 1 & -1 \\ 
  1 & 1 & -1 & 1 & -1 &  -1  & && 1 & 1 & -1 & 1 & -1 & -1 \\
  -1 & 1 & 1 & 1 & -1&   0  & &&-1 & 1 & 1 & -1 & 1 & -1 \\ 
  1 & 1 & 1 & -1 & 1 & -1  & &&1 & -1 & 1 & -1 & 1 & -1 \\ 
 -1 & -1 & -1 & -1 & -1 & 1   &&& 1 & 1 & -1 & -1 & 1 & 1 \\ 
1 & -1 & 1 &  -1 & -1 &  1   &&& -1 &  -1 &  -1 & 1    & 1 &  1 \\
 
   -1 & -1 & -1 & -1 & -1 & -1  & &&  -1 & -1 & -1 & -1 & 1 & -1 \\ 
 
  -1 &  1 & -1 & 1 & 1 & 1  &&&  -1 & -1 & 1 & -1 & -1 & 1 \\
 
 1 & -1 &  1 & 1 &  1 & 1   &&&  -1 & 1 & -1 & -1 & -1 & 1 \\
  
\end{tabular}
\end{table}

\begin{table}[ht]
\caption{$n = 15, \ k = 6$ Bayesian $A_s$- and $D_s$-optimal designs for main effect models with potential two-factor interaction effects. The Bayesian-$A_s$ comes from $15 \leq \tau^{-2} < 100$ while the Bayesian-$D_s$ has $20 \leq \tau^{-2} \leq 100$.}\label{tab:Bayes6F15R}
\centering

\begin{tabular}{rrrrrr rr rrrrrr}

\multicolumn{6}{c}{\text{Bayesian $A_s$-optimal}} &&& \multicolumn{6}{c}{\text{Bayesian $D_s$-optimal}} \\

   -1 &  1 & 1 & -1 & 0 & 0 & &&  1 &  1  & 1 &  1 &  1 & -1 \\
   
   -1 & -1 &  1 &  1 &  1 &  1&&& 1 &  -1  & 1 & -1 & 1 & 1  \\ 
  
  -1 &  1 & -1 & 1 & 1 & 1  &&& -1 & 1 & 1 & -1 & 1 & 1 \\  
  
   1 & -1 & -1 & 1 & -1 &  1  &&&      -1 & -1 & 1 &  1 &  1 & -1 \\ 
  
   -1 &  1 & -1 & 1 & -1 & -1 &&&   1 & -1 &  -1 & 1 &  1 & -1  \\ 
   
  -1 & -1 & -1 & -1 & 1 & -1  &&& -1 & -1 & -1 & -1 & -1 &  1  \\ 
   
   1 & -1 & -1 &  1 &  1 & -1 &&&  -1 & -1 &1 & -1 & -1 & -1  \\
   
   -1 & -1 & -1 & -1 & -1 &  1 &&&    -1 & 1 & -1 & 1 & -1 & -1 \\ 
   
 1 & -1 &  1 & -1 &  1 &  1&  &&  1 &1 & -1 & 1 & 1 &  1 \\ 
   
   -1 & -1 & 1 & 1 & -1 & -1 &&&  -1 & -1 &-1 & 1 &  1 &  1 \\ 
   
   1 &  1 &  1 & 1 & -1 & 1 &&&  1 & 1 &-1 & -1 & -1 & -1  \\ 
   
    1 &  1 & -1 & -1 & -1 & -1 &&&   1 & 1 &1 & -1 & -1 & 1  \\ 
   
  1 & -1 & 1 & -1 & -1 & -1 &&&  1  &  -1 & 1  &  1 &  -1 &  1 \\
  
   1 & 1 & -1 & -1 & 1 & 1 &&&   -1 &1&  -1 & -1 & 1 & -1 \\ 
   
   1 & 1 & 1 &  1 & 1 & -1 &&&  -1 & 1& 1 & 1 & -1 & 1 \\ 

\end{tabular}
\end{table}

\begin{table}[ht]
\caption{$n = 15, \ k = 6$ Bayesian $A_s$-optimal design for main effect models with potential two-factor interaction effects with $\tau^{-2} = 10$. \label{tab:BayesAnonInt}}
\centering

\begin{tabular}{rrrrrr}

\multicolumn{6}{c}{\text{Bayesian-}$A_s$}  \\

1 & 1 & -1 & -1 & 1 & 1 \\ 
 -1 & 1 & -1 & 1 & -1 & 1 \\ 
1 & -1 & -0.62 & 0.62 & -1 & 1 \\ 
1 & 1 & -1 & -1 & -1 & -1 \\ 
1 & -1 & 1 & -1 & 0.64 & -0.64 \\ 
1 & 1 & 1 & 1 & 1 & 1 \\ 
-1 & -1 & 1 & 1 & -1 & -1 \\ 
-0.60 & 0.60 & 1 & -1 & -1 & 1 \\ 
 -1 & 1 & 1 & -1 & 1 & -1 \\ 
-1 & -1 & -1 & -1 & 1 & 1 \\ 
 -1 & -1 & 1 & 1 & 1 & 1 \\ 
 1 & 1 & 1 & 1 & -1 & -1 \\ 
1 & -1 & -1 & 1 & 1 & -1 \\ 
 -1 & 1 & -1 & 1 & 1 & -1 \\ 
 -1 & -1 & -1 & -1 & -1 & -1 \\ 

\end{tabular}
\end{table}

\clearpage




\subsection{Section 4.3 Designs and Results for $k=8, 10$}

\begin{figure}[h]
\centering
\includegraphics[width=0.9\textwidth]{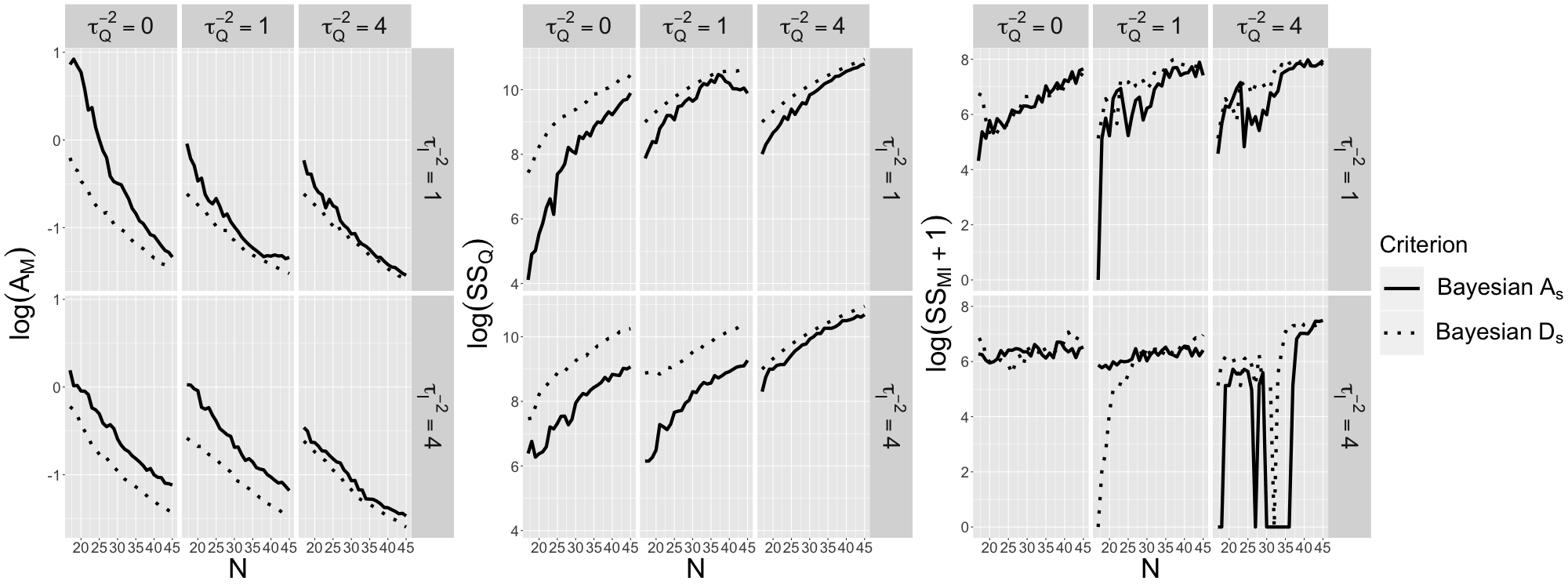}
\caption{Performance measures for the Bayesian $D_s$-optimal and Bayesian $A_s$-optimal designs when $k = 8$ found with $\tau_Q^{-2} \in \{0,1,16\}$ and $\tau_{I}^{-2} \in \{1,16\}$. (Left) The $A_S$-criterion for the main effect model on the log scale. (Middle) The sum of squares of the off-diagonals for the quadratic terms on the log scale. (Right) The sum of squares of the cross products of the main effects and interactions on the log scale with offset $1$. }\label{fig:BayesRSMk8}
\end{figure}

\begin{figure}[h]
\centering
\includegraphics[width=0.9\textwidth]{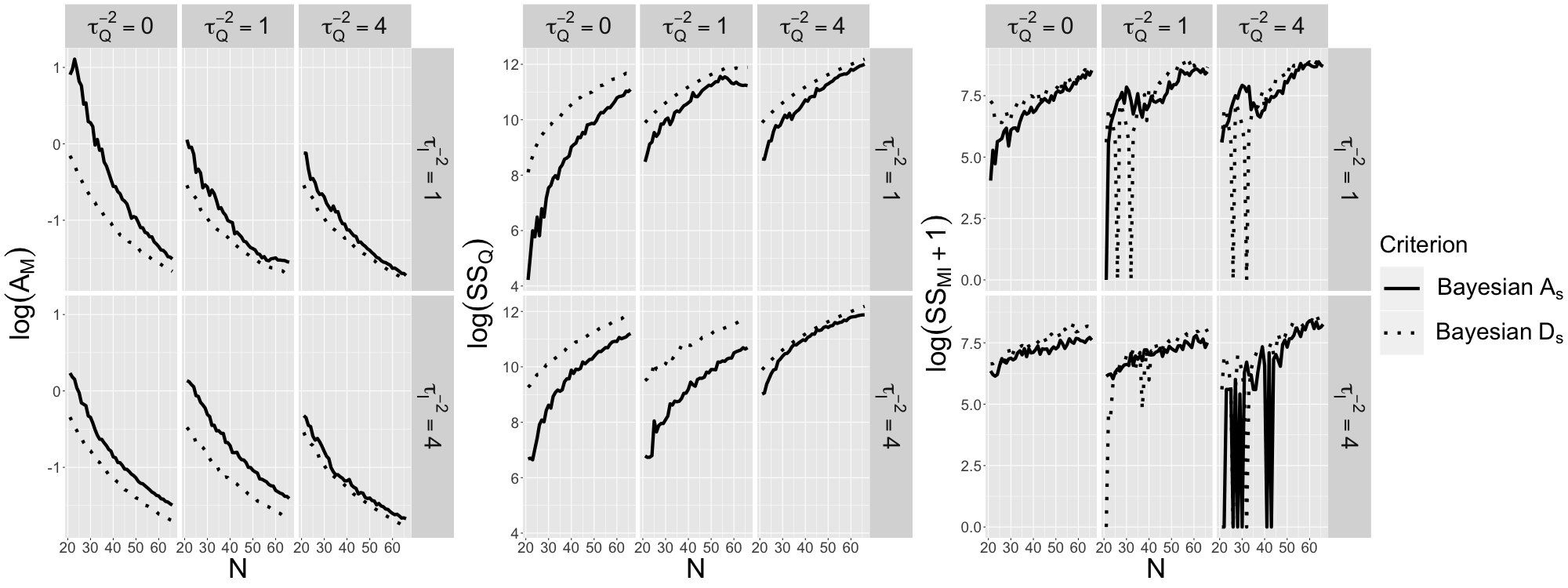}
\caption{Performance measures for the Bayesian $D_s$-optimal and Bayesian $A_s$-optimal designs when $k = 10$ found with $\tau_Q^{-2} \in \{0,1,16\}$ and $\tau_{I}^{-2} \in \{1,16\}$. (Left) The $A$-value for the first-order model on the log scale. (Middle) The sum of squares of the off-diagonals for the quadratic terms on the log scale. (Right) The sum of squares of the cross products of the main effects and interactions on the log scale with offset $1$.  }\label{fig:BayesRSMk10}
\end{figure}

Results for factors $k = 8$ and $k = 10$ are fairly analogous to the $k = 6$ case discussed in Section~4.3 of the main paper. The results, found in Figures~\ref{fig:BayesRSMk8} and \ref{fig:BayesRSMk10}, continue to indicate that the Bayesian $A_s$-optimal design prioritizes estimation of the quadratic effects, while the Bayesian $D_s$-optimal design does not. For $\tau_Q^{-2} = \tau_I^{-2} = 16$, the Bayesian-$A$ for both $k = 8$ and $k = 10$ produced even more designs with $SS_{MI} = 0$, while the Bayesian $D_s$-optimal design for $k = 8$ found no such designs, and for $k = 10$ found few relative to the Bayesian-$A$. This adds to the previous conclusion that the Bayesian $A_s$-optimal criterion reduces aliasing for designs compared with the Bayesian $D_s$-optimal.

\clearpage

\bibliographystyle{asa}
\bibliography{Aopt}

\end{document}